\documentclass[10pt,letter,envcountsame]{article}

\usepackage{amsmath,amstext,amsgen,latexsym}
\usepackage{amstext,amssymb,amsfonts,latexsym}
\usepackage{theorem}
\usepackage{pifont}

 \newcommand{\bs}{\bigskip}
 \newcommand{\ms}{\medskip}
 \newcommand{\n}{\noindent}
 \newcommand{\s}{\smallskip}
 \newcommand{\hs}[1]{\hspace*{ #1 mm}}
 \newcommand{\vs}[1]{\vspace*{ #1 mm}}

 \newcommand{\setempty}{\mathrm{\O}}
 \newcommand{\real}{\mathbb{R}}
 
 \newcommand{\nat}{\mathbb{N}}
 \newcommand{\integer}{\mathbb{Z}}
 \newcommand{\rational}{\mathbb{Q}}
 \newcommand{\complex}{\mathbb{C}}
 
 \newcommand{\field}{\mathbb{F}}

 \newcommand{\prob}{{\mathrm{Prob}}}

 \newcommand{\ie}{\textrm{i.e.},\hspace*{2mm}}
 \newcommand{\eg}{\textrm{e.g.},\hspace*{2mm}}

 \newcommand{\AAA}{{\cal A}}
 \newcommand{\BB}{{\cal B}}

 \newcommand{\DD}{{\cal D}}

 \newcommand{\PP}{{\cal P}}
 \newcommand{\OO}{{\cal O}}
 \newcommand{\TT}{{\cal T}}
 \newcommand{\UU}{{\cal U}}

 \newcommand{\p}{\mathrm{P}}
 \newcommand{\np}{\mathrm{NP}}

 \newcommand{\bqp}{\mathrm{BQP}}

\def\bbox{\vrule height6pt width6pt depth1pt}
\theoremstyle{plain}
\theoremheaderfont{\bfseries}
\newtheorem{theorem}{Theorem}
\newtheorem{lemma}[theorem]{Lemma}
\newtheorem{proposition}[theorem]{Proposition}
\newtheorem{corollary}[theorem]{Corollary}

{\theorembodyfont{\rmfamily}
     \newtheorem{definition}[theorem]{Definition}}
{\theorembodyfont{\rmfamily} }
{\theorembodyfont{\rmfamily} }

\newenvironment{proof}{\par \noindent
            {\bf Proof. \hs{2}}}{\hfill$\Box$ \vspace*{3mm}}

\newenvironment{proofof}[1]{\vspace*{5mm} \par \noindent
         {\bf Proof of #1.\hs{2}}}{\hfill$\Box$ \vspace*{3mm}}

 \newcommand{\ceilings}[1]{\lceil #1 \rceil}
 \newcommand{\floors}[1]{\lfloor #1 \rfloor}

 \newcommand{\ignore}[1]{}

 \newcommand{\qubit}[1]{| #1 \rangle}
 \newcommand{\bra}[1]{\langle #1 |}
 \newcommand{\braket}[2]{\langle #1 | #2 \rangle}

\newcommand{\Pre}{\mathrm{Pre}}

\newcommand{\bfzero}{{\bf 0}}

\begin{document}
%%%%%%%%%%%%%%%%%
\pagestyle{plain}
\setcounter{page}{1}

\begin{center}
{\Large {\bf Quantum List Decoding of Classical Block Codes of \s\\
Polynomially Small Rate from Quantumly Corrupted Codewords}}\footnote{An early version appeared in the Proceedings of the 13th Computing: The Australasian Theory Symposium (CATS 2007), pp. 153--162, Ballarat, Australia, January 30--February 2, 2007. This work was in part supported by the Mazda Foundation.} \bs\bs\\

{Tomoyuki Yamakami}\footnote{Current Affiliation: 
Faculty of Engineering, University of Fukui, 3-9-1 Bunkyo, 
Fukui, 910-8507, Japan} \bs\ms\\
\end{center}

%%%%%%%%%%%%%%%%%%%%%%%%%%%%%%%%%%%%%%%%%%%%
\begin{quote}
\n{\bf Abstract.} \hs{1} Given a classical error-correcting block code, the task of quantum list decoding is to produce from any quantumly corrupted codeword a short list containing all messages whose codewords exhibit high ``presence'' in the quantumly corrupted codeword.
Efficient quantum list decoders have been used to prove a quantum hardcore property of classical codes.
However, the code rates of all known families of efficiently quantum list-decodable codes are, unfortunately, too small for other practical applications. To improve those known code rates, we
prove that a specific code family of polynomially small code rate
over a fixed code alphabet, obtained by concatenating generalized Reed-Solomon codes as outer codes with Hadamard codes as inner codes,
has an efficient quantum list-decoding algorithm if its codewords  have relatively high codeword presence in a given quantumly corrupted codeword. As an immediate application, we use the quantum list decodability of this   code family to solve a certain form of quantum search problems in polynomial time.
When the codeword presence becomes  smaller, in contrast, we show that the quantum list decodability of generalized Reed-Solomon codes with high confidence is closely related to the efficient solvability of the following two problems: the noisy polynomial interpolation problem and the bounded distance vector problem.
Moreover, assuming that $\np\nsubseteq \bqp$, we also prove that  no efficient quantum list decoder exists for the generalized Reed-Solomon codes.

\vspace{.1in}

\end{quote}

%%%%%%%%%%%%%%%%%%%%%%%%%%%%%%%%%%%%%%%%%%%%%%
%%%%%%%%%%   APPENDIX   %%%%%%%%%%%%%%%%%%%%%%
%%%%%%%%%%%%%%%%%%%%%%%%%%%%%%%%%%%%%%%%%%%%%%
%%%%%%%%%%%%%%%%%%%%%%%%%%%%%%%%%%%%
\sloppy
\section{Quantum List Decoding}

Classical list decoding, whose notion is attributed to Elias \cite{Eli57} and Wozencraft \cite{Woz58} in late 1950s, has recently drawn significant attention after Sudan's \cite{Sud97} discovery of an efficient list-decoding algorithm for well-studied Reed-Solomon codes beyond its ``traditional'' error-correction radius. List decoding has since then found useful applications to cryptography as well as computational complexity theory
(see, \eg survey articles of Sudan \cite{Sud00} and Trevisan \cite{Tre04}). For a wider range of  applications to, in particular,  quantum computations, an introduction of quantum analogue of such list decoding is an inevitable consequence.

In a seminal paper of Kawachi and Yamakami \cite{KY06} (following an early work of Adcock and Cleve \cite{AC02} on biased oracles) published first in 2006, a notion of \emph{quantum list decoding} of classical block codes arose quite naturally in their study of {\em quantum hardcore functions} for arbitrary (strongly) quantum one-way functions. A goal of quantum list decoding in Kawachi and Yamakami's {\em implicit-input explicit-output model} is to produce a relatively short list of message candidates by means of oracle queries to a faulty quantum encoding procedure given as a form of {\em oracle}. This model of quantum list decoding slightly differs from a conventional transmission model between a sender and a receiver through a noisy channel, particularly,  in the following aspects.
Given an original message hidden to the receiver, assumed is the existence of
a faulty quantum encoding procedure (called as a {\em quantum-computationally corrupted codeword} or {\em quantumly corrupted codeword}) that tries to generate a code symbol at each specified block location of a desired codeword induced from the original message.
 To recover the hidden message from this quantumly corrupted codeword, the receiver is  allowed to  access the quantumly corrupted codeword {\em repeatedly}, partly because he cannot duplicate ``unknown'' quantum states by a quantum-mechanical principle.
The quantumly corrupted codeword is likely to behave   ``adversarially'' and hinder the receiver's effort of recovering uniquely the original  message. Quite often, however, it is sufficient to produce a reasonably short list of message candidates including all the messages whose corresponding codewords are in close proximity to the given quantumly corrupted codeword, and thus this list certainly contains the hidden message. This ``closeness'' is scaled by the notion of {\em codeword presence} (or {\em presence}, in short), which indicates the average   probability of  obtaining successfully each block symbol of the target codeword from the quantumly corrupted codeword (see Kawachi and Yamakami \cite{KY06} for an intuition behind this notion).
Because of these differences, the classical list decodability does not generally imply the quantum list decodability.
To construct hardcore functions is the primary purpose of quantum list decoding by Kawachi and Yamakami \cite{KY06}, and their study of quantum list decoding was centered at a natural question of what types of classical block codes are efficiently quantum list decodable.

In the past literature showed several families of block codes that are classical/quantum list decodable in polynomial time. The first of such examples is a family of {\em Hadamard codes}.
In the case of
classical list decoding, Goldreich and Levin \cite{GL89} proved the classical list decodability of the binary Hadamard codes, and subsequently Goldreich, Rubinfeld, and Sudan \cite{GRS95} presented a general list-decoding algorithm for the $q$-ary Hadamard codes. Concerning quantum list decoding, by contrast, Adcock and Cleve \cite{AC02} essentially proved that the binary Hadamard codes are quantum list decodable in polynomial time. For the $q$-ary Hadamard codes, a fast quantum list-decoding algorithm was given by Kawachi and Yamakami \cite{KY06}. They also presented two additional quantum list-decodable codes: {\em shifted Legendre symbol codes} and {\em pairwise equality codes}.
A common feature of these codes is that
they all have exponentially small code rate, where the {\em rate} of a code is a ratio of message length (or a dimension of the code) and codeword's block length. For instance, the rate of the binary Hadamard code is exactly $n/2^n$ for message length $n$.
Notice that, in a practical setting, code rate and block length are important factors in designing error-correcting codes.
In particular, a family of polynomial-time classical list-decodable codes of polynomially small rate over the binary code alphabet finds numerous applications in the fields of cryptography and computational complexity theory (refer to, \eg survey articles by Sudan \cite{Sud00} or Trevisan \cite{Tre04}).

All known efficiently quantum list-decodable code families have so far {\em exponentially small} code rate, which is extremely smaller than the code rates of many practical codes. It is therefore natural to ask whether there exists an efficiently quantum list-decodable code of polynomially small rate and of fixed alphabet size for any given bias parameter.
This paper is profoundly motivated by this intriguing question and,
as its main theorem, it will successfully prove the existence of such a code family; more strongly, we will show the following statement.
\begin{theorem}{\rm [Main Theorem]}\label{c-q-list-decoding}
Let $q$ be any prime constant. For any constant $k\geq1$, there exist a polynomially-time computable function $t:\nat\rightarrow\nat^{+}$ and a classical block $(t(n),n)_{q}$-code family $C$  such that
\renewcommand{\labelitemi}{$\circ$}
\begin{enumerate}\vs{-1}
  \setlength{\topsep}{-2mm}%
  \setlength{\itemsep}{0mm}%
  \setlength{\parskip}{0cm}%

\item $C$ is polynomial-time classically list decodable with confidence $5/6$, and

\item $C$ is polynomial-time quantumly list decodable with presence at least $1/q+1/n^k$ and confidence $2/3$.
\end{enumerate}\vs{-1}
This code family $C$ has code rate $n/t(n)$, which is only polynomially small.
\end{theorem}
The rest of this paper is dedicated to proving this theorem and seeking its application.

To obtain the desired code family stated in the main theorem,
we will initially seek a well-studied code family. A family of {\em generalized Reed-Solomon (GRS) codes} has relatively large code rate; however,
it usually has large alphabet size.
{}From this code family, we will build a family of codes of high code rate over a fixed code alphabet by an idea of Forney \cite{For66}.
In this paper,  we will use in Section \ref{sec:codes} a concatenated code $C^{GRS\mbox{-}H}$ of Guruswami and Sudan \cite{GS00}, which is obtained by  concatenating the generalized Reed-Solomon codes with the Hadamard codes.
Our key claim---Theorem \ref{high-rate}---states that the codes $C^{GRS\mbox{-}H}$ (with an adequate choice of code parameters) are efficiently quantum list decodable as far as their codeword presence is relatively high.  Theorem \ref{c-q-list-decoding} follows immediately from this claim, because $C^{GRS\mbox{-}H}$ was already proven to be classically list decodable (Guruswami and Sudan \cite{GS00}).
As the first step toward the proof of Theorem \ref{high-rate}, we will demonstrate in Proposition \ref{concatenated_code} that this concatenated code family possesses efficient quantum list decodability, provided that the generalized Reed-Solomon codes are efficiently quantum list decodable. This claim will be proven in Section \ref{sec:quantum-reduction} by employing a technique of constructing an efficient ``quantum reduction'' between two quantumly corrupted codewords.
An advantage of this proof technique is that it requires no {\em soft information}, which is a key ingredient in the classical case of Guruswami and Sudan \cite{GS99,GS00}.

Our next step is to show in Lemma \ref{Reed-Solomon} that the generalized Reed-Solomon codes are indeed efficiently quantum list decodable,
by partially applying a {\em polynomial reconstruction algorithm} of Guruswami and Sudan \cite{GS99}, as far as  a target codeword has relatively high presence in a given quantumly corrupted codeword. Unfortunately, the use of such a classical algorithm  makes the query complexity of our quantum list decoder quite high.
On the contrary, as the presence becomes lower, it seems to become harder to solve efficiently the quantum list-decoding problem.
For instance, when the presence is arbitrary close to a reciprocal of the code alphabet size, we can convert an efficient quantum list-decoding algorithm to an efficient quantum algorithm that even solves a certain $\np$-complete problem. This immediately leads to  an unlikely consequence that every $\np$-problem can be solved efficiently on a quantum computer with high success probability. In a similar vein, we will present a direct connection between quantum list decodability of the generalized Reed-Solomon codes and the quantum solvability of two classical problems: the {\em noisy polynomial interpolation problem} (NPIP) of Naor and Pinkas \cite{NP99} and the {\em bounded distance vector problem} (BDVP), both of which will be defined in Sections \ref{sec:NPIP}--\ref{sec:BDVP}.
To be more precise, we will show that (1) if the generalized Reed-Solomon codes are quantumly list decodable, then the NPIP is quantumly solvable and (2) if the BDVP is quantumly solvable, then the generalized Reed-Solomon codes are quantumly list decodable.

Our quantum list-decoding algorithm for the aforementioned concatenated code finds an immediate application to certain types of problems.  Our example in this paper is an {\em NBQP-search problem}, in which, given a polynomial-time quantum algorithm and an input instance, we want to find a classical witness of polynomial size that forces the algorithm to accept the input with high probability. We will show in Section \ref{sec:search-problem} that solving this search problem on average implies solving it in worst case. This can be compared to a classical case of an $\np$-search problem of Kumar and Sivakumar \cite{KS99}.

In line of the study on quantum list decoding, we will make a brief discussion in Section \ref{sec:local-list-decoding} on another notion of {\em local quantum list decoding} based on an \emph{implicit-input implicit-output model} where an outcome of a list-decoding algorithm is a list of {\em descriptions} of quantum-circuit list decoders rather than a list of messages. Similarly to the classical case of Sudan, Trevisan, and Vadhan \cite{STV01}, we can apply our quantum list decoder for generalized Reed-Solomon codes to conduct local quantum list decoding for the Reed-M{\"u}ller codes. As an immediate consequence, we can prove the so-called {\em hardness amplification} of quantum circuits, following the argument of Sudan, Trevisan, and Vadhan \cite{STV01}.

%%%%%%%%%%%%%%%%%%%%%%%%%%%%%%%%%%%%%%%
\section{Foundations of Quantum List Decoding}

This section explains basic notions and notation concerning quantum list decoding. Throughout this paper, let $\nat$ denote the set of all  {\em natural numbers} (\ie nonnegative integers) and set $\nat^{+}=\nat-\{0\}$. For any positive integers $m$ and $n$ with $m\leq n$, the notation $[m,n]_{\integer}$ means the integer set $\{m,m+1,m+2,\ldots,n\}$ and $[n]$ is  the shorthand for $[1,n]_{\integer}$ whenever $n\geq1$. For any number $q\in\nat^{+}$,  $\field_{q}$ (or $GF(q)$) denotes a {\em finite (Galois) field} of size $q$. When $q$ is a prime number, we often express  the elements of $\field_{q}$ in terms of the numbers in $[0,q-1]_{\integer}$. We sometimes use a prime power $q^m$ rather than a prime $q$. Conventionally, we also identify each vector in $(\field_{q})^m$ with its corresponding element in $\field_{q^m}$. Let $\rational$ and $\complex$ respectively denote the sets of all \emph{rational numbers} and of all \emph{complex numbers}. We further set $\rational^{\geq0}=\{r\in\rational\mid r\geq0\}$.

For a finite alphabet $\Sigma$, a {\em string} $x$ over $\Sigma$ is a finite sequence of symbols from $\Sigma$, and $|x|$ denotes the {\em length} of $x$ (\ie the number of all the occurrences of symbols in $x$).

%%%%%%%
\subsection{Classical Block Codes}\label{sec:block-code}

We briefly explain classical block (error-correcting) codes, which are key objects of our interest. Roughly speaking,
a {\em (block) code} is a set of strings of the same length over
a finite alphabet $\Sigma$ and each string of a code is indexed by a message and is called a {\em codeword}. In this paper, we mostly deal with a {\em family of codes}, each of which corresponds to a different message length $n$ in $\nat$.
Such a code family can be specified in general by a series $(\Sigma_n,I_n,\Gamma_n)$ of triplets composed of {\em message space} $\Sigma_n$, {\em index set} $I_n$, and {\em code alphabet} $\Gamma_n$ for each {\em message length}\footnote{This parameter is also known as the {\em dimension} or {\em information length} of a code.} $n$ (which serves as a ``basis parameter'' in this paper).

As standard nowadays in computational complexity theory, we view a code $C$  (or $C^{(n)}$, to emphasize ``$n$'')
for each fixed message length $n$ as a ``function'' that maps
$\Sigma_n\times I_n$ to $\Gamma_n$. For convenience, let the {\em code size}  $N(n) = |\Sigma_n|$  and let the {\em code alphabet size}  $q(n)=|\Gamma_n|$. It is also convenient to assume that $\Sigma_n =  (\Sigma')^n$ for a certain fixed {\em message alphabet}  $\Sigma'$ so that $n$ actually represents the {\em length} of messages in $\Sigma_n$ over $\Sigma'$; in this case, $n= \log_{|\Sigma'|}N(n)$ holds for every length $n\in\nat$. For instance, if $\Sigma' = \{0,1\}$, then all messages can be expressed in binary. By abbreviating $C(x,y)$ as $C_{x}(y)$, we treat $C_{x}(\cdot)$ as a function mapping $I_n$ to $\Gamma_n$ and we call it a {\em codeword}, whose {\em block length} (or {\em code length}) $M(n)$ equals $|I_n|$. Since the elements in $I_n$ serve as indices of block locations of a codeword, it is often assumed that $I_n=\{0,1,\ldots,M(n)-1\}$ so that each element of $I_n$ can be expressed in $\ceilings{\log_2M(n)}$ bits. For convenience, we also identify $C_x$ with the vector $(C_{x}(0),C_{x}(1),\cdots,
C_{x}(M(n)-1))$ in the {\em ambient space} $(\Gamma_n)^{M(n)}$ of dimension $M(n)$. Because we mainly work on a finite field, we often regard $\Gamma_n$ as a finite field $\field_{q(n)}$  of order $q(n)$.

The {\em rate} of a code $C$ is defined to be the ratio $n/M(n)$. The {\em (Hamming) distance} $d(C_x,C_y)$ between two codewords $C_x$ and $C_y$ is the number of non-zero components in the vector $C_x - C_y$. The {\em minimal distance} $d(C^{(n)})$ (or $d(n)$, in short) of the codes of message length $n$ is
the smallest distance between any pair of distinct codewords associated with the messages of length $n$.
In contrast, $\Delta(C_x,C_y)$ denotes the {\em relative (Hamming) distance} $d(C_x,C_y)/M(n)$. The above-described code is simply called an  $(M(n),n)_{q(n)}$-code\footnote{The reader should be aware that, in some literature, the notation $(M(n),\Gamma_n)_{q(n)}$ is used instead.} (or $(M(n),n,d(n))_{q(n)}$-code, to emphasize the minimal distance $d(n)$ of the code of message length $n$).
For readability, we often drop a length parameter $n$ from both subscripts and argument places whenever we discuss
a set of codewords of a ``fixed'' message length $n$. A {\em linear $(M(n),n)_{q(n)}$-code} forms a $n$-dimensional vector space in $\left(\field_{q(n)}\right)^{M(n)}$.

\ms
\n{\sf Hadamard Codes $\mathrm{HAD}$.} Let $n$ be any message length used as a parameter, and let $q$ be any prime number. A {\em $q$-ary Hadamard code family} $\mathrm{HAD}^{(q)}=\{\mathrm{HAD}^{(q,n)}\}_{n\in\nat}$ consists of all $(q^n,n,q^n-q^{n-1})_q$-codes $\mathrm{HAD}^{(q,n)}:(\field_{q})^n\times(\field_q)^n\rightarrow \field_q$ obtained as follows. For each  message $x=(x_1,x_2,\ldots,x_n)$ in $(\field_q)^n$,  $\mathrm{HAD}^{(q,n)}(x,r)$ equals $\sum_{i=1}^{n}x_i r_i\;\mathrm{mod}\;q$, where $r=(r_1,r_2,\ldots,r_n)$ is in the index set $(\field_q)^n$.

\ms
\n{\sf (Normalized) Generalized Reed-Solomon Codes $\mathrm{GRS}$.} Let $q$ be any prime number and let $k$ and $n$ be any two positive integers satisfying that $n\leq k\leq q$. A {\em (normalized) generalized Reed-Solomon code family} $\mathrm{GRS} = \{\mathrm{GRS}^{(k,n,q)}\}_{n,k\in\nat}$ consists of all $(k,n,k-n+1)_q$-codes defined as follows. Let $x=(x_1,x_2,\ldots,x_n)\in(\field_q)^n$ be any message and let $D_k$ be a fixed set of $k$ distinct elements (called {\em code locators}) in $\field_q$. Let $\mathrm{GRS}^{(k,n,q)}:(\field_q)^n\times D_k\rightarrow \field_q$ be defined as $\mathrm{GRS}^{(k,n,q)}(x,r) = \sum_{i=1}^{n}x_i r^{i-1} \;\mathrm{mod}\;q$, which is a polynomial of degree at most $n-1$ with $r\in D_k$. Occasionally, we expand the domain $D_k$ of $\mathrm{GRS}^{(k,n,q)}_x$ to the entire field $\field_{q}$.

%%%%%%%%%%
\subsection{Quantumly Corrupted Codewords and Codeword Presence}\label{sec:quantumly-corrected}

A {\em quantum bit} (or a {\em qubit}, in short) is a unit vector in the complex space $\complex^2$, and a {\em quantum state} is generally a tensor product of some of these qubits. To express such a quantum state, we customarily use Dirac's notation. For instance, a quantum state $\qubit{\phi}$ of two qubits can be expressed as $\qubit{\phi_1}\otimes\qubit{\phi_1}$, where $\qubit{\phi_1}$ and $\qubit{\phi_2}$ are both qubits; however, we often abbreviate $\qubit{\phi_1}\otimes\qubit{\phi_2}$ as $\qubit{\phi_1}\qubit{\phi_2}$.   An execution of a \emph{quantum algorithm} on an input instance corresponds to a series of applications of unitary operations, and it is usually modeled by a ``computation'' of a \emph{quantum Turing machine} (Bernstein and Vazirani \cite{BV97}; Yamakami \cite{Yam99,Yam03}) or a \emph{quantum circuit} (Yao \cite{Yao93}). We use the notation $\AAA(x)$ (or more formally, $\AAA\qubit{x}$) to denote a quantum state obtained after executing quantum algorithm $\AAA$ on classical
input $x$ (which is formally given in the form of quantum state $\qubit{x}$).
When we refer to an {\em output} of $\AAA$ on $x$, we mean a classical string that is obtained by \emph{measuring} (or \emph{observing}) the quantum state $\AAA(x)$ in the standard computational basis, where a measurement is a projection onto a certain Hilbert space. For simplicity, we say that a quantum algorithm {\em runs in polynomial time} if its corresponding quantum Turing machine halts within time polynomial in the length of each  input. Similar to the complexity classes $\p$ and $\np$, $\bqp$ denotes the collection of all (classical) decision problems that can be solved by quantum algorithms in polynomial time with success probability at least $2/3$.
 For more details on quantum computation, the reader may refer to, \eg Nielsen and Chuang \cite{NC00}.

\s

Let us consider a quantum procedure that tries to encode a classical message into its codeword. In general, a quantum computation tends to interact with an  outside system of a currently operating quantum system, causing a quantum corruption of the computation.
Hence, our process of quantum encoding may be corrupted. A corrupted process of such quantum encoding can be described as an application of a certain form of unitary operator. As noted before, when $q(n)$ is a prime number, we represent each element in $\field_{q(n)}$ as an integer in $[0,q(n)-1]_{\integer}$, which is further expressed in binary.
In their 2006 conference paper, Kawachi and Yamakami coined the terminology---a {\em quantum-computationally corrupted codeword} or {\em  quantumly corrupted codeword}---to describe such a unitary operator $O$, with two fixed parameter functions $\ell(n)$ and $m(n)$ mapping $\nat$ to $\nat$, that satisfies the following condition: for any two strings $r\in I_{n}$ and  $s\in\{0,1\}^{m(n)}$ and any number $\ell(n)$, there exists a quantum state $\qubit{\phi_{r,z}}$ of $\ell(n)$ qubits such that
\begin{equation}\label{quantumly-corrupted-codeword}
O\qubit{r}\qubit{s}\qubit{0^{\ell(n)}} = \sum_{z\in \{0,1\}^{m(n)}}\alpha_{r,z}\qubit{r}\qubit{s\oplus z}\qubit{\phi_{r,z}},
\end{equation}
where the notation $\oplus$ denotes the bitwise XOR,
$\qubit{\phi_{r,z}}$ indicates garbage information produced when we apply
the operator $O$ to the three registers, and the amplitudes $\{\alpha_{r,z}\}_{r,z}$ satisfy that $\sum_{z\in\{0,1\}^{m(n)}}|\alpha_{r,z}|^2=1$ for every index $r\in I_n$.
Since $O$ is a unitary operator, so is its {\em inverse} $O^{-1}$.
Another important notion of Kawachi and Yamakami is ``codeword
presence'' in $O$.
The {\em presence} of codeword $C_x$ in $O$, denoted $\Pre_{O}(C_x)$, is the average probability of obtaining the correct values $C_x(r)$ by a measurement over all indices $r\in I_n$; namely, $\Pre_{O}(C_x) = (1/M(n))\sum_{r\in I_n}|\alpha_{r,C_x(r)}|^2$.

%%%%%%%%%%%%%%%%%%%%
\subsection{Asymptotic Behaviors of Codeword Presence}\label{sec:presence-distance}

The value of \emph{codeword presence} is a key to the performance of a quantum list decoder.  We will briefly argue asymptotic behaviors of codeword presence for arbitrary quantumly corrupted codewords in a fashion similar to classical cases of Guruswami, H{\aa}stad, Sudan, and Zuckerman \cite{GHSZ02}.
For this purpose, we need to expand the existing notions of presence and (Hamming) distance of codewords in a more general fashion. Notice that these generalized presence and distance are applied only to this subsection.

Let $n$ be any message length and define $W_n$ to be the set of all vectors $w = (w_{r,z})_{r\in I_n,z\in\field_{q(n)}}\in [0,1]^{q(n)M(n)}$ (where each $w_{r,z}$ may be viewed as the probability $|\alpha_{r,z}|^2$ of obtaining $(r,z)$ after  measuring a quantumly corrupted codeword) satisfying the restriction that $\sum_{z\in[0,q(n)-1]_{\integer}}w_{r,z}=1$ for each index $r\in[0,M(n)-1]_{\integer}$, where $M(n)=|I_n|$.
 For every $w\in W_n$, it follows that  $\sum_{r}\sum_{z}w_{r,z} = M(n)$.
Next, we consider the set $V_n$ of all codewords (viewed as a vector) $a=(a_{r})_{r\in I_n}\in ([0,q(n)-1]_{\integer})^{M(n)}$. We embed each codeword $a$ into $W_n$ by the special mapping $v$, defined as $v(a)= (\delta^{(a)}_{r,z})_{r\in I_n,z\in\field_{q(n)}}\in\{0,1\}^{q(n)M(n)}$, where $\delta^{(a)}_{r,z}$ is $1$ if $a(r)=z$, and $0$ otherwise. Moreover, for any code (seen as a subset of $V_n$) $C^{(n)}$, let $v(C^{(n)})=\{v(a)\mid a\in C^{(n)}\}$.
Obviously, $v(V_n)\subseteq W_n$ holds.

Using the above notations, let us generalize the notions of distance and presence as follows.
For any pair $v,w\in W_n$, we define $d(v,w) = M(n) - \braket{v}{w}$, where
$\braket{\cdot}{\cdot}$ denotes the standard {\em inner product}.
This generalized notion naturally expands the standard notion of the distance $d(\cdot,\cdot)$ because, for any $a,b\in V_n$, we have
\[
d(v(a),v(b)) = M(n)-\braket{v(a)}{v(b)}) = M(n) - |\{(r,z)\mid a(r) = b(r) =z \}| = d(a,b).
\]
Moreover, for any two vectors $a\in V_n$ and $w\in W_n$, define $\Pre_{w}(a) = \frac{1}{M(n)}\braket{v(a)}{w}$. We then obtain
\[
\Pre_{w}(a) = \frac{M(n)-d(v(a),w)}{M(n)} = \frac{\braket{v(a)}{w}}{M(n)}
= 1 - \frac{d(v(a),w)}{M(n)}.
\]

%%%%%%%%
%%%%%%%%

First, we wish to obtain an \emph{asymptotic lower bound} of codeword presence in terms of minimal relative distance $\lambda$. For this purpose, we will introduce the notation $QL^{poly}(\lambda)$ for the minimal possible ``presence'' $\varepsilon$, with which, for an arbitrary family of block codes with minimal relative distance $\lambda$, the cardinality of all messages having codeword presence of at least $\varepsilon$ is polynomially bounded.
More precisely, let $C=\{C^{(n)}\}_{n\in\nat}$ be any $(M(n),n,d(n))_{q(n)}$-code family and let $\Delta(C^{(n)}) = d(C^{(n)})/M(n)$ express the relative distance of $C^{(n)}$. For each pair $w\in W_n$ and $\varepsilon\in[0,1]$, we write $E(w,\varepsilon)$ for the set $\{a\in V_n \mid \Pre_{w}(a)\geq \varepsilon\}$.
For any function $f:\nat\rightarrow\nat$ and any number $n\in\nat$, the notation $presence(C,f)(n)$ denotes $\min\{\varepsilon\in\real^{\geq 0} \mid \forall w\in W_n [\,|E(w,\varepsilon)\cap C^{(n)}|\leq f(n)\,]\}$
and we  set $Pre(C,f)= \limsup_{n\rightarrow\infty}\left\{\frac{presence(C,f)(n)}{M(n)}\right\}$.
In addition, let $QL_{f}(\lambda) = \inf_{C:\Delta(C)\geq \lambda}\{Pre(C,f)\}$, where   $\Delta(C)=\liminf_{n\rightarrow\infty}\{\Delta(C^{(n)})\}$. For each fixed  constant $c\in\nat$, we set $QL^{poly}_c(\lambda) = \sup_{a>0}\{QL_{f_a^{(c)}}(\lambda)\}$, where $f_a^{(c)}(n)=an^c$ for any number  $n\in\nat$. Finally,   $QL^{poly}(\lambda)$ is set to be  $\limsup_{c\rightarrow\infty}\{QL^{poly}_c(\lambda)\}$.

\begin{proposition}\label{lower-bound-presence}
\sloppy Let $c$ be any positive constant and let $\lambda$ be any number in $[0,1]$, representing a minimal relative distance. It holds that  either   $QL^{poly}_c(\lambda) \geq 1/q + (1-1/q)\left(1-\lambda/(1-1/q) + \lambda/an^c(1-1/q)\right)^{1/2}$ or $QL^{poly}_c(\lambda) \geq 1/q +
(1-1/q)\left(1-\lambda/(1-1/q) \right)^{1/2}$. Therefore, $QL^{poly}(\lambda) \geq 1/q + (1-1/q)\left(1-\lambda(1-1/q)\right)^{1/2}$ follows.
\end{proposition}

In certain extreme cases, it holds that $QL^{poly}(0) = 1$ and $QL^{poly}(1) \geq   \sqrt{1/q} + (1-\sqrt{1/q})/q$.
It remains open whether the equality $QL^{poly}(\lambda) = 1/q + (1-1/q)\left(1-\lambda(1-1/q)\right)^{1/2}$ holds or not.

%%%%%%%%%%%%%%%%%%%%%%%%%%%
%%%%%%%%%%%%%%%%%%%%%%%%%%%

Next, we will show an \emph{asymptotic upper bound} of codeword presence, particularly, in terms of the rate of a ``linear'' $(M(n),n,d(n))_{q(n)}$-code family $C=\{C^{(n)}\}_{n\in\nat}$. For convenience, we write $rate(C^{(n)})$ for the code rate $n/M(n)$ of $C^{(n)}$. Here, let $R$ be any code rate in $[0,1]$ and let $f:\nat\rightarrow\nat$ be any function. We
define $QU_f(R) = \sup_{C:rate(C)\geq R}\{Pre(C,f)\}$, where $rate(C)= \liminf_{n\rightarrow\infty} \{rate(C^{(n)})\}$. With this notation, for each   constant $c>0$, we write $QU^{const}_c(R)$ for  $QU_{f_c}(R)$, where $f_c$ is a constant function defined as $f_c(n)=c$ for all numbers $n\in\nat$.
Define $QU^{const}(R)$ to be $\lim\sup_{c\rightarrow\infty} \{QU^{const}_{c}(R)\}$.

\begin{proposition}\label{QU-upper-bound}
Fix an odd prime number $q$. For every constant $c\in\nat^{+}$ with $c>2(q-1)$ and every code rate $R\in(0,1)$, it holds that $QU^{const}_c(R)\geq 1 - q^{-\frac{(1+2R)c-q}{(q-2)c}}$. Therefore, $QU^{const}(R) \geq 1- q^{-\frac{1+2R}{q-2}}$ follows.
\end{proposition}

For readability, we place the proofs of Propositions \ref{lower-bound-presence}--\ref{QU-upper-bound}
in Appendix.

%%%%%%%%%%%%%%%%%%%%%%%%%%%
%%%%%%%%%%%%%%%%%%%%%%%%%%%
\subsection{Kawachi-Yamakami Implicit-Input Explicit-Output Model}\label{sec:input-output-model}

To formulate the notion of quantum list decoding, this paper deals with a specific model in which we
 {\em implicitly} take  a quantumly corrupted codeword as a form of ``oracle'' and then we output a list of messages {\em explicitly} after accessing the oracle by way of oracle queries.
A process of making an oracle query and then receiving its oracle answer is conventionally assumed to take a unit time.  Upon this {\em implicit-input explicit-output model}, the {\em quantum list-decoding problem} (QLDP) for a classical block code family $C$ can be described as follows. First, let $C=\{C^{(n)}\}_{n\in\nat}$ be any $(M(n),n,d(n))_{q(n)}$-code family with message space $\Sigma_n$ and let $\OO$ be any set of quantumly corrupted codewords for $C$. Taking a {\em bias parameter} $\varepsilon:\nat\to[0,1]$, we define the $\varepsilon$-QLDP as:

\ms
\n{\sc $\varepsilon$-Quantum List Decoding Problem ($\varepsilon$-QLDP) for Code Family $C$ with respect to $\OO$}
\renewcommand{\labelitemi}{$\circ$}
\begin{itemize}\vs{-2}
  \setlength{\topsep}{-2mm}%
  \setlength{\itemsep}{0mm}%
  \setlength{\parskip}{0cm}%

\item {\sc Input:} a message length $n$ and a value $1/\varepsilon(n)>0$.

\item {\sc Implicit Input:} an oracle $O\in\OO$ representing a quantumly corrupted codeword for $C^{(n)}$.

\item {\sc Output:} a list of messages including all messages $x\in \Sigma_n$ that satisfy the inequality
$\Pre_{O}(C_x) \geq 1/q(n) + \varepsilon(n)$. For convenience, we refer to such a list as a {\em valid list} for the $\varepsilon$-QLDP.
\end{itemize}

Our goal is to solve the problem $\varepsilon$-QLDP for $C$  using an efficient quantum algorithm that makes an oracle access to a given quantumly corrupted codeword in $\OO$  with success probability at least $\delta(n)$, which is given as a {\em confidence parameter}. Here, let us formally introduce the notion of a {\em quantum list-decoding algorithm} (or simply, a {\em quantum list decoder}) that works with two parameters: bias $\varepsilon$ and confidence $\delta$.

\begin{definition}[quantum list decoding]
Let $C$ be any code family, let $\varepsilon(n)$ be any bias parameter, and let $\delta(n)$ be any confidence parameter. A {\em quantum list-decoding algorithm} (or a {\em quantum list decoder}) for $C$ with bias $\varepsilon$ and confidence $\delta$ is a quantum algorithm $\AAA$ that solves the $\varepsilon$-QLDP for $C$ with success probability at least $\delta(n)$.  If $\AAA$ further runs in time polynomial in
$(n,1/\varepsilon(n),1/\delta(n))$,
it is called a {\em polynomial-time quantum list-decoding algorithm}
for $C$.
\end{definition}

The {\em list size} of a quantum list decoder with respect to  input size $n$ refers to the maximal size of any valid list produced by the algorithm on any input of size $n$.
In certain applications, the list size of a single valid list plays a crucial role; for instance, when a quantum list decoder produces only a single valid list $L$ (along all measured outcomes) with probability at least $\delta(n)$, certain ``advice'' of size $\ceilings{\log_{|\Sigma|}|L|}$ over a message alphabet $\Sigma$ may help  specify a hidden message $x$ {\em uniquely} with the same success probability.

\s

A close connection between quantum list decoding and (strongly) quantum one-way functions was exhibited by Kawachi and Yamakami \cite{KY06}. The rest of this subsection briefly discusses a further relationship between quantum list decoding and  a restricted form of quantum one-way functions, called  {\em quantum super one-way functions}, which can be seen as a natural extension of quantum one-way permutations.

\begin{definition}[quantum super one-wayness]\label{super-oneway}
Let $f$ be any function mapping $\Sigma^*$ to $\Sigma^*$ with length function $\ell:\nat\rightarrow\nat$, that is, $|f(x)|=\ell(|x|)$ for every $x$. This function $f$ is called {\em quantum super one-way} if (i) there exists a polynomial-time quantum algorithm $\AAA$ such that, for every input $x$ of length $n$,  $\AAA\qubit{x}\qubit{0^{\ell(n)}}\qubit{0^{e(n)}}  = \qubit{x}\qubit{f(x)}\qubit{\phi_{x}}$ holds for a certain unit-norm quantum state $\qubit{\phi_{x}}$ of $e(n)$ qubits and (ii) for any positive polynomial $p$ and any polynomial-time quantum algorithm $\BB$, the probability that $\BB$ on input $\qubit{1^n}\qubit{f(x)}\qubit{\phi_{x}}$
outputs $x$ of length $n$ is at most $1/p(n)$ for all but finitely many strings $x$.
\end{definition}

In comparison with Definition \ref{super-oneway}, the quantum one-wayness formulated by Kawachi and Yamakami \cite{KY06} requires that $\BB\qubit{f(x)}$ outputs $x$ only with negligible probability whereby the information $\qubit{\phi_x}$ is {\em hidden} from the adversary $\BB$ who tries to invert $f$. Definition \ref{super-oneway}, on the contrary, indicates that $\BB$ cannot output $x$  with non-negligible probability even though $\qubit{\phi_x}$ is given to $\BB$ besides $f(x)$ as supplemental information.
In computational cryptography, this notion naturally arises. A typical example of super one-way function is a {\em quantum one-way permutation} obtained by replacing further the quantum state $\qubit{\phi_x}$ in Definition \ref{super-oneway} with $\qubit{0^m}$, which is obtained, for example, by uncomputing a deterministic procedure that computes $f(x)$ from $x$.

In what follows, for any index $i$, the notation $(f(x))_{i}$ denotes the $i$th bit of the value $f(x)$ whenever $1\leq i\leq |f(x)|$.

\begin{lemma}
Let $f$ be any quantum super one-way function with its {\em length function} $m(n)\in n^{O(1)}$ (\ie $|f(x)|=m(|x|)$). Consider an   $(m(n),n,d(n))_{q(n)}$-code $C$ whose codeword $C_x(r)$ is $(f(x))_{r}$. For every positive polynomial $p$, this code $C$ cannot be polynomial-time quantum list decodable with confidence $1/p(n)$.
\end{lemma}

\begin{proof}
Let $f$ be a quantum super one-way function with its length function
$m(n)$, where $m(n)$ is polynomially bounded, and consider an   $(m(n),n,d(n))_{q(n)}$-code $C$ satisfying  $C_x(r) = (f(x))_{r}$ for any $x$ and $r$. Since a certain polynomial-time quantum algorithm must compute $f$ exactly as stated in Definition \ref{super-oneway}, by modifying this algorithm slightly, we obtain another
polynomial-time quantum algorithm, say, $\AAA$ that computes $C(x,r)$.
Without loss of generality, we may assume that, for every $n\in\nat^{+}$, every $x\in\Sigma^n$, and every $r\in[\ell(n)]$,
$\AAA\qubit{x}\qubit{r}\qubit{0}\qubit{0^{e(n)}} =\qubit{x}\qubit{r}\qubit{C_x(r)}\qubit{\phi_{x}}$ holds for a certain quantum state $\qubit{\phi_{x}}$ that depends only on $x$. Here, we fix $x$ of length $n$ arbitrarily and define
$
O_x\qubit{r}\qubit{s}\qubit{0^{e(n)}} =
\qubit{r}\qubit{s\oplus C_x(r)}\qubit{\phi_{x}}
$
for any strings $r$ and $s$. Notice that $\Pre_{O_x}(C_x)=1$ holds. Toward a contradiction, assume that $C$ has a polynomial-time quantum list decoder $\BB$ such that,  since the presence of $C_x$ in $O_x$ is $1$, $\BB$  on input $1^n$  produces the hidden string $x$  with probability at least $1/p(n)$  for a certain fixed positive polynomial $p$, where ``$1^n$'' indicates an input representing ``$n$'' in the definition of the $\varepsilon$-QLDP. We want to invert $f$ in polynomial time. For this goal,  we define a quantum algorithm $\DD$ as follows.
\begin{quote}\vs{-1}
On input $\qubit{1^n}\qubit{f(x)}\qubit{\phi_x}$, where $n=|x|$, we run the quantum list decoder $\BB$ on input $1^n$ using $O_x$ as an oracle.  However, whenever $\BB$ makes an oracle query $\qubit{r}\qubit{s}\qubit{t}$ to the oracle $O_x$, we simulate the behavior of $O_x$ as follows. We  generate an oracle answer $\qubit{r}\qubit{s\oplus (f(x))_r}\qubit{\phi_x}$ directly using the input information. Finally, we output an outcome of $\BB$. Since $\Pre_{O_x}(C_x)=1$, the outcome of $\BB$ must be $x$ itself.
\end{quote}
The above algorithm $\DD$ thus inverts $f$ correctly with probability at least $1/p(n)$. This implies that $f$ cannot be quantum super one-way, a contradiction against our assumption. Therefore, $C$ is not polynomial-time quantum list decodable with confidence $1/p(n)$.
\end{proof}

%%%%%%%%%%%%%%%%%%%%%%%%%%%%%%%%
\section{Codes of Polynomially Small Rate}\label{sec:codes}

The proof of our main theorem (Theorem \ref{c-q-list-decoding}) requires a suitable code family of polynomially small code rate over a fixed code alphabet. Such a code family can be obtained by Forney's  \cite{For66} idea of concatenating two appropriate code families.
In Section \ref{sec:concatenation}, we will claim that this concatenated code family is efficiently quantumly list decodable for a certain choice of code parameters. This claim---Theorem \ref{high-rate}---then leads to the   main theorem. Therefore, our primary goal is to conduct necessary ground work that  leads to the proof of Theorem \ref{high-rate}. For the sake of readability, we will split the proof into two claims---Proposition \ref{concatenated_code} and Lemma \ref{Reed-Solomon}---and this section will prove only the proposition, leaving the lemma to Section \ref{sec:Reed-Solomon}.  A key proof technique of this section in handling the concatenated code is a {\em quantum reduction} between two quantumly corrupted codewords, maintaining ``similar'' codeword presence values.

%%%%%%%%
\subsection{Concatenated Codes}\label{sec:concatenation}

A typical way to build a family of classical block codes that have desired code rate and desired code alphabet size is to compose two appropriate block codes with certain necessary code properties. This is
Forney's \cite{For66} novel method of creating so-called {\em concatenated codes}. In our case, concatenating an appropriate generalized Reed-Solomon code with its  matching Hadamard code, we can build a code of polynomially small code rate and constant code alphabet size. For such a code family, we will prove its efficient quantum list decodability, provided that
the generalized Reed-Solomon codes have efficient quantum list decoders.

More formally, let us consider two block codes $C_1$ and $C_2$ such that $C_1$ is an $(M_1,n_1,d_1)_{q^{n_2}}$-code and $C_2$ is an $(M_2,n_2,d_2)_{q}$-code. Let $x=(x_1,x_2,\ldots,x_{n_1})$ be any message of length $n_1$, where each entry $x_i$ is taken from $\Sigma^{n_2}$ over  a $q$-letter alphabet $\Sigma$. Since $x_i$ can be expressed as an $n_2$-letter string, $x$ can be viewed as a string of total length $n_1n_2$ over $\Sigma$.  By taking
the inner code $C_2$ concatenated with the outer code $C_1$, the concatenated code $C=C_2\odot C_1$  is defined as $C(x,r,s) = C_2(C_1(x,r),s)$ for every triplet $(x,r,s)$. This code $C$ becomes an $(M_1M_2,n_1n_2,d)_{q}$-code with $d$ satisfying $d\geq d_1d_2$, where $d_1d_2$ is called the {\em design distance}.

For our purpose of this section, we choose the concatenated code $C^{GRS\mbox{-}H}[n,q,\theta]$ given by Guruswami and Sudan \cite{GS00}.
This concatenated code is obtained from a certain generalized Reed-Solomon code used as an outer code together with an appropriate Hadamard code used as an inner code.  Following Guruswami and Sudan \cite{GS00}, here we choose three parameters $(n,q,\theta)$ with $n,q\in\nat$ and $\theta\in[0,1]$ that satisfy  $n\geq1$, $q\geq2$, $n = m q^m\theta$, and $q^m\theta\in\nat$ for a certain number $m\in\nat$. In what follows, we freely identify elements in $(\field_q)^n$ with elements in $\field_{q^n}$ in the standard fashion.

%%%%%%%%%%%%
\ms
\n{\sf Concatenated Code $C^{GRS\mbox{-}H}[n,q,\theta]$.}
The concatenated code  $C^{GRS\mbox{-}H}[n,q,\theta]$ is defined by $C^{GRS\mbox{-}H}[n,q,\theta] =  \mathrm{HAD}^{(q,m)} \odot  \mathrm{GRS}^{(q^m,q^m\theta,q^m)}$. This is a $(q^{2m},n,d)_q$-code, where $n=mq^m\theta$ and  $d\geq (1-1/q)(1-\theta)q^{2m}$ (design distance).
{}From $n=mq^m\theta$, we obtain $\log{n} = \log{mq^m\theta}$, from which  $m= \frac{\log{n}-\log{m}+\log(1/\theta)}{\log{q}}$ follows. This implies $\frac{\log(1/\theta)}{\log q} \leq m \leq n$; thus, $q^{m} = \frac{n}{m\theta} \leq \frac{n\log{q}}{\theta\log(1/\theta)}$.
As long as $q$ is fixed and $\theta=\Omega(1/n^k)$ holds for a certain constant $k\in\nat^{+}$, $q^{m}$ is upper-bounded by $O(n^{k+1}/\log{n})$.
Hence, the code rate $n/q^{2m}$ is lower-bounded by $\frac{c\log{n}}{n^{k}}$ for a certain constant $c>0$.

\ms

This concatenated code family $C^{GRS\mbox{-}H} = \{C^{GRS\mbox{-}H}[n,q,\theta]\}_{n,q,\theta}$ is proven by Guruswami and Sudan \cite{GS00} to be efficiently classically list decodable; that is, there exists a polynomial-time probabilistic algorithm that produces, from any classically corrupted codeword (or conventionally, a \emph{received word}) $w$, a list containing all messages $x$ whose codewords are all at distance close
to $w$. To prove Theorem \ref{c-q-list-decoding}, it therefore suffices to show that the code family $C^{GRS\mbox{-}H}$ is also  quantumly list decodable in an efficient manner for appropriately chosen  parameters. In a more general fashion, we intend to show the following statement. Let $\TT$ denote the collection of all tuples $(n,m,q,\theta)$ such that $m,q,\in\nat^{+}$, $q\geq2$, $\theta\in[0,1]$, $q^m\theta\in\nat$, and $n=mq^m\theta$.

\begin{theorem}\label{high-rate}
For each $n\in\nat^{+}$, assume that a parameter tuple  $(m,q,\theta,\varepsilon,\delta)$ satisfies the following conditions: $(n,m,q,\theta)\in\TT$, $\varepsilon,\delta\in[0,1]$,
$2(1-1/q)^2(1/M+\varepsilon')<\varepsilon^2$, and $2(1-1/q)^2(1-1/M-\varepsilon')\sqrt{\frac{\Delta}{1+M\varepsilon'}}<\varepsilon^2$
for a certain $\varepsilon'\in(0,1)$, where $M= q^m$ and $\Delta = 2(n-1)\log(M^2/(1-\delta))$.
The concatenated code family $C^{GRS\mbox{-}H} = \{C^{GRS\mbox{-}H}[n,q,\theta]\}_{n,q,\theta}$ with the above conditions has a quantum list decoder with bias $\varepsilon$ and confidence $\delta$ running in time polynomial in $(n,q,1/\varepsilon,1/\delta,1/(1-\delta))$.
\end{theorem}

Here, we give the proof of Theorem \ref{c-q-list-decoding} using Theorem \ref{high-rate}. Although it is possible to relax the conditions stated in Theorem \ref{high-rate} further, they are sufficient to prove Theorem \ref{c-q-list-decoding}.

\begin{proofof}{Theorem \ref{c-q-list-decoding}}
Fix a prime number $q$, a constant $k\in\nat^{+}$, and a confidence parameter $\delta$. Here, we set $\varepsilon = 1/n^k$. We also choose other parameters $(n,m,q,\theta)\in\TT$ and $M = q^m = O(n^{\ell})\cap\Omega(n^{8k+4})$ for a certain fixed
constant $\ell\geq 8k+4$ and consider the code $C^{GRS\mbox{-}H}[n,q,\theta]$.
In this case, it holds that $\theta = n/mM = \Omega(1/n^{\ell-1})$ since $m=\log{M}/\log{q}$. This guarantees the polynomially small code
rate of $C$.
Let us define $t(n)=M^2$. Note that
the  value $\Delta = 2(n-1)\log(M^2/(1-\delta))$ satisfies $\Delta= O(n\log{n}) =O(n^{1.4})$. For simplicity, set $\alpha = 1/M+\varepsilon'$. Now, defining
$\varepsilon' = \sqrt{1/M}$, we obtain $M\alpha = 1+\sqrt{M} = \Omega(n^{4k+2})$. It thus follows that $2(1-\alpha)\sqrt{\frac{\Delta}{M\alpha}}= O(n^{0.7}/n^{2k+1})$. Since $\varepsilon=1/n^k$, we obtain  $2(1-\alpha)\sqrt{\frac{\Delta}{M\alpha}}< \varepsilon^2$ and $2\alpha<\varepsilon^2$ for any sufficiently large $n$. Since all premises of Theorem \ref{high-rate} are fulfilled, there must exist  a quantum list decoder with bias $\varepsilon$ and confidence $\delta$. This quantum list decoder runs in time polynomial in $n$ since the parameters $(q,\delta)$ are constants. This completes the proof.
\end{proofof}

Let us return to Theorem \ref{high-rate}. This theorem, in fact, follows from two technical claims: Proposition \ref{concatenated_code} and Lemma \ref{Reed-Solomon}.
We will prove in Proposition \ref{concatenated_code} that the concatenated code family $C^{GRS\mbox{-}H} = \{C^{GRS\mbox{-}H}[n,q,\theta]\}_{n,q,\theta}$ has a polynomial-time quantum list decoder for an appropriate choice of three  parameters $(n,q,\theta)$, assuming that the generalized Reed-Solomon codes are quantum list decodable in polynomial time.
This last assumption will be later eliminated, in Lemma \ref{Reed-Solomon},  completing the proof of Theorem \ref{high-rate}.

\begin{proposition}\label{concatenated_code}
For each $n\in\nat^{+}$, let $(q,\theta,m,\varepsilon,\varepsilon',\delta)$ satisfy the following conditions: $(n,m,q,\theta)\in\TT$,  $\varepsilon,\varepsilon',\delta\in[0,1]$,
and $\varepsilon^2\geq (1-1/q)^2(1/M+\varepsilon')$. If the $(M,M\theta,(1-\theta)M+1)_{M}$-generalized Reed-Solomon code has a quantum list decoder with bias $\varepsilon'$ and confidence $\delta$ running in time polynomial in $(n,q,1/\varepsilon',1/\delta,1/\theta)$, where $M= q^m$, then $C^{GRS\mbox{-}H}[n,q,\theta]$ has a quantum list decoder with bias $\varepsilon$ and confidence $\delta$ running in time polynomial in $(n,q,1/\varepsilon',1/\delta,1/\theta)$.
\end{proposition}

Note that the confidence $\delta$ for the GRS-code in Proposition \ref{concatenated_code} is carried over to the confidence for the concatenated code $C^{GRS\mbox{-}H}[n,q,\theta]$. The proposition is an important ingredient of Theorem \ref{high-rate} and its proof will be given in the subsequent subsection.

%%%%%%
\subsection{A Quantum Reduction Technique}\label{sec:quantum-reduction}

Aiming at proving Proposition  \ref{concatenated_code}, we wish to construct a ``quantum reduction'' between two quantumly corrupted codewords. Such a reduction, say, from $O$ to $O'$ can be described as a quantum algorithm that, on input of the form $\qubit{r}\qubit{s}\qubit{t}$, computes the outcome $O'\qubit{r}\qubit{s}\qubit{t}$ by invoking a number of oracle calls to $O$ as well as $O^{-1}$. This can be seen as a strong form of well-known {\em Turing reduction} between two languages.

Here, let $C$ be any $(q^m,n/m)_{q^m}$-code, which is, as before, treated as a function $C(x,r)$ mapping from $(\field_{q})^{\frac{n}{m}}\times \field_{q^m}$ to $\field_{q^m}$ whenever $n/m\in\nat^{+}$. Recall that we freely identify $(\field_{q})^m$ with $\field_{q^m}$. As a technical lemma essential for the proof of Proposition \ref{concatenated_code}, we will show a general result concerning a concatenated code $D = \mathrm{HAD}^{(q,m)}\odot C$. Since the $q$-ary Hadamard code $\mathrm{HAD}^{(q,m)}$ is used as an inner code, we can rephrase $D$ as
\[
D(x,r,s) = C(x,r) \cdot s \;\;\mathrm{mod}\;q
\]
for any $r,s\in \field_{q^m}$ and any $x\in(\field_{q^m})^n$.

In what follows, let us aim at constructing a quantum reduction between quantumly corrupted codewords $O_C$ and $O_D$ associated with the codes $C$ and $D$, respectively.
For convenience, we introduce new terminology. For any unitary transform $U$, we  say that a quantum algorithm $\AAA$ {\em realizes} $U$ if, for any {\em basis} quantum  state $\qubit{r}$, $\AAA$ on input $\qubit{r}$ exactly produces the quantum state  $U\qubit{r}$.
This notion will help describe a quantum reduction from $O_C$ to $O_D$.

\begin{lemma}\label{from-C-to-D}
Let $C$ and $D$ be the codes given as above.  For any quantumly corrupted codeword $O_{C}$ for $C$, there exist a polynomial-time quantum algorithm $\AAA$ and a quantumly corrupted codeword $O_D$ for $D$ such that
\renewcommand{\labelitemi}{$\circ$}
\begin{enumerate}\vs{-2}
  \setlength{\topsep}{-2mm}%
  \setlength{\itemsep}{0mm}%
  \setlength{\parskip}{0cm}%

\item $\Pre_{O_D}(D_x) = 1/q + \left( 1 - 1/q \right)\Pre_{O_C}(C_x)$; and
\item $\AAA$ realizes $O_D$ with one oracle access to $O_C$.
\end{enumerate}
\end{lemma}

For the proof of Proposition \ref{concatenated_code}, we need to weaken the notions of ``quantumly corrupted codeword'' and ``realization.''
A {\em generalized quantumly corrupted codeword} $O$ is defined by
Eq.(\ref{quantumly-corrupted-codeword}) except that we require only the inequality $\sum_{z}|\alpha_{r,z}|^2\leq1$ among the amplitudes $\{\alpha_{r,z}\}_{r,z}$. The codeword presence of $C_x$ in each of the operators $O_k$ is defined as before. Let $\OO=\{O_k\}_{k\in[q-1]}$ denote a series of generalized quantumly corrupted codewords.
For this series $\OO$, we also define  the \emph{average (codeword) presence}  $av\Pre_{\OO}(C_x)$ of $C_x$ in $\OO$ to be $(1/(q-1))\sum_{k\in[q-1]}\Pre_{O_k}(C_x)$.
For a series $\UU=\{U_k\}_{k\in[q-1]}$ of unitary operations, we say that a quantum algorithm $\AAA$ {\em weakly realizes} $\UU$ if $\AAA$ on input $\qubit{k}\qubit{r}\qubit{0}$ generates a certain quantum state and, after tracing out the third register by the observable $\qubit{0}$, it becomes  $\qubit{k}\otimes U_k\qubit{r}$.

\begin{lemma}\label{from-D-to-C}
Let $C$ and $D$ be the codes given as above. For any quantumly corrupted codeword $O_{D}$ for $D$, then there exist a polynomial-time quantum algorithm $\AAA$ and a series $\OO=\{O_k\}_{k\in[q-1]}$ of generalized quantumly corrupted codewords for $C$ such that
\renewcommand{\labelitemi}{$\circ$}
\begin{enumerate}\vs{-2}
  \setlength{\topsep}{-2mm}%
  \setlength{\itemsep}{0mm}%
  \setlength{\parskip}{0cm}%

\item $av\Pre_{\OO}(C_x) \geq  (q/(q-1))^2(\Pre_{O_D}(D_x) - 1/q)^2$;
 and
\item $\AAA$ weakly realizes $\OO$ with one oracle access to each of $O_D$ and  $O_{D}^{-1}$.
\end{enumerate}
\end{lemma}

Lemma \ref{from-D-to-C} gives a fast quantum reduction from $O_D$ to $O_C$.
{}From this lemma directly follows Proposition \ref{concatenated_code}. Before proving Lemmas \ref{from-C-to-D}--\ref{from-D-to-C}, we briefly describe the proof of the proposition.

%%%%%%%
\begin{proofof}{Proposition \ref{concatenated_code}}
Let $n\in\nat^{+}$ be any length parameter and assume that all other parameters $(m,q,\theta,\varepsilon,\varepsilon',\delta)$ satisfy the premise of the
proposition. Hereafter, we set $M=q^{m}$ and $D = C^{GRS\mbox{-}H}[n,q,\theta]$ for brevity.
Let us assume that the $(M,M\theta,(1-\theta)M+1)_{M}$-generalized Reed-Solomon code has a polynomial-time quantum list decoder, say, $\AAA$ with bias $\varepsilon'$ and confidence $\delta$.
Take any quantumly corrupted codeword $O$ for $D$. Our goal here is to find from $O$ all messages $x$ that satisfy the inequality $\Pre_{O}(C_x)\geq 1/q+\varepsilon$ in time polynomial in $(n,q,1/\varepsilon',1/\delta,1/\theta)$ with confidence $\delta$.

Since $D = \mathrm{HAD}^{(q,m)}\odot \mathrm{GRS}^{(M,M\theta,M)}$,  Lemma \ref{from-D-to-C} helps reduce $O$ to a series $\OO'=\{O'_k\}_{k\in[q-1]}$ of generalized quantumly corrupted codewords for the outer code $\mathrm{GRS}^{(M,M\theta,M)}$ so that $\OO'$ can be weakly realized by a certain polynomial-time quantum algorithm, say, $\BB$  with the following average presence condition:
\begin{eqnarray*}
av\Pre_{\OO'}\left(\mathrm{GRS}^{(M,M\theta,M)}_x\right)
&\geq& \left(\frac{q}{q-1}\right)^2 \left( \Pre_{O}(D_x) -
\frac{1}{q}\right)^2 \\
&\geq& \left(\frac{q}{q-1}\right)^2\varepsilon^2 \;\; \geq\;\;
\frac{1}{M}+\varepsilon',
\end{eqnarray*}
where the last inequality follows directly from the bound $\varepsilon^2\geq (1-1/q)^2(1/M+\varepsilon')$, which is given as a part of the premise of the proposition.
In other words, the average value of $\Pre_{O'_k}(\mathrm{GRS}^{(M,M\theta,M)}_x)$ over all $k\in[q-1]$ is lower-bounded by $1/M+\varepsilon'$. Thus, we can choose an index $k_0\in[q-1]$ for which $\Pre_{O'_k}(\mathrm{GRS}^{(M,M\theta,M)}_x)\geq 1/M+\varepsilon'$.
By our assumption, for this $k_0$, $\AAA$ correctly produces a list including all messages $x$ satisfying $\Pre_{O'_k}(\mathrm{GRS}^{(M,M\theta,M)}_x)\geq 1/M+\varepsilon'$ with confidence $\delta$.

Let us consider the following quantum algorithm, which uses $\AAA$ and $\BB$ as subroutines.

\begin{quote}
On input, we first set $k=0$ and, by incrementing $k$ by one, we inductively run the quantum list decoder $\AAA$ with $O'_k$ as an oracle to produce a list of message candidates. During inductive steps, we always  append new candidates to the existing list. Whenever a query is made, we run $\BB$ to generate its oracle answer. This is possible because $\BB$ weakly realizes $\OO'$. Eventually, we reach $k_0$ and we then obtain a list  containing of all messages $x$ satisfying  $\Pre_{O'}(\mathrm{GRS}^{(M,M\theta,M)}_x)\geq  1/M+\varepsilon'$ with probability at least $\delta$.
\end{quote}

This algorithm is obviously a quantum list decoder and it produces with confidence $\delta$ a list that contains all messages $x$ satisfying $\Pre_{O}(D_x) \geq 1/q+\varepsilon$. This completes the proof.
\end{proofof}

Next, we want to prove Lemmas \ref{from-C-to-D}--\ref{from-D-to-C}. We begin with the proof of Lemma \ref{from-D-to-C}.

\begin{proofof}{Lemma \ref{from-D-to-C}}
Let $C$ be any $(q^m,n/m)_{q^m}$-code. We denote by $D$ the concatenated code $\mathrm{HAD}^{(q,m)}\odot  C$ and
assume that $O_D$ satisfies $O_D\qubit{r,s}\qubit{u}\qubit{0^{\ell(n)}} = \sum_{z\in\field_q}\alpha_{r,s,z}\qubit{r,s}\qubit{u\oplus z}\qubit{\phi_{r,s,z}}$  for any $r,s\in \field_{q^m}$, where $\ell(n)$ indicates the size of garbage information $\qubit{\phi_{r,s,z}}$. Note that $\Pre_{O_D}(D_x) = q^{-2m}\sum_{r,s\in\field_{q^m}}|\alpha_{r,s,D_x(r,s)}|^2$ and that $\sum_{z\in\field_{q}}|\alpha_{r,s,z}|^2 =1$ for every pair $(r,s)$.

We wish to define the desired quantum algorithm $\AAA$ and the desired series $\OO=\{O_k\}_{k}$ of generalized quantumly corrupted codewords that can be weakly realized by $\AAA$ using $O_D$ as an oracle. To describe the algorithm $\AAA$, we utilize  a special unitary transform $U$ over $[q-1]$ acting as  $U\qubit{0} = (1/\sqrt{q-1})\sum_{k\in[q-1]}\qubit{k}$ as well as  a \emph{quantum Fourier transform} $F_{q}$ over $\field_{q}$ that acts as  $F_{q}\qubit{s} = q^{-1/2} \sum_{w\in \field_{q}}\omega_{q}^{s\cdot w}\qubit{w}$ for any  $s\in \field_{q}$. It was proven by van Dam, Hallgren, and Ip \cite{DHI06} that $F_{q}$ can be approximated to within error $\eta$ on a quantum computer in time polynomial in $(\log{q},\log(1/\eta))$.

%%%%%%%%%%%%%%%%%%%%%%%%%%%%%%%%%%%
\vspace*{5mm}
\hrule\vspace*{2mm}
\n{\sc Quantum Algorithm $\AAA$:}
\renewcommand{\labelitemi}{$\circ$}
\begin{enumerate}\vs{0}
  \setlength{\topsep}{-2mm}%
  \setlength{\itemsep}{0mm}%
  \setlength{\parskip}{0cm}%\begin{enumerate}
%%%%%%%%%%%%%%%%%%%%%%%%%%%%%%%%%%%%

\item[(1)] Start with an initial quantum state $\qubit{\psi_1}= \qubit{k}\qubit{r}\qubit{0^m}\qubit{0}\qubit{0^{\ell}}$.

\item[(2)] By applying the quantum Fourier transform $(F_{q})^m$ to the third register, we generate the quantum state $\qubit{\psi_2} = q^{-m/2}\sum_{s\in(\field_q)^m}\qubit{k}\qubit{r,s}\qubit{0}\qubit{0^{\ell}}$, where $\qubit{r,s}$ is a shorthand for $\qubit{r}\qubit{s}$.

\item[(3)] \sloppy Apply $O_D$ to the last three registers. This step transforms the quantum state $\qubit{\psi_2}$ into
$
\qubit{\psi_3} = q^{-m/2}\sum_{s\in(\field_q)^m}\sum_{z\in\field_{q}}\alpha_{r,s,z}\qubit{k}\qubit{r,s}\qubit{z}\qubit{\phi_{r,s,z}}.
$

\item[(4)] \sloppy Apply the {\em phase encoding} of Kawachi and Yamakami \cite{KY06}; that is, encode the content of the fourth register into the ``phase'' together with the information on $k$ to obtain
$
\qubit{\psi_4} = q^{-m/2} \sum_{s\in(\field_q)^m}\sum_{z\in\field_q} \omega_q^{k\cdot z} \alpha_{r,s,z} \qubit{k}\qubit{r,s}\qubit{z}\qubit{\phi_{r,s,z}}.
$

\item[(5)] Apply $O_{D}^{-1}$, the {\em inverse} of $O_D$, to the last four registers. The resulted state $\qubit{\psi_5}$ can be expressed as
$
\sum_{s\in(\field-q)^m}\sum_{z\in\field_q} \beta_{k,r,s,z}\qubit{k}\qubit{r,s}\qubit{0}\qubit{0^{\ell}} + \qubit{k}\qubit{\Delta_{k,r}}
$
with certain amplitudes $\beta_{k,r,s,z}$ and a certain vector $\qubit{\Delta_{k,r}}$ whose last two registers does not contain the term  $\qubit{0}\qubit{0^{\ell}}$.
Each amplitude $\beta_{k,r,s,z}$ is calculated as
\[
\beta_{k,r,s,z} = \bra{k}\bra{r}\bra{s}\bra{0}\bra{0^{\ell}}  I\otimes O_D^{-1}\qubit{\psi_k}
=  \frac{1}{q^{m/2}}\omega_q^{k\cdot z}|\alpha_{r,s,z}|^2,
\]
where $I$ is the identity transform.
The quantum state $\qubit{\psi_5}$ is thus written in the form
\[
\frac{1}{q^{m/2}} \sum_{s\in(\field_q)^m}\sum_{z\in\field_q} \omega_q^{k\cdot z} |\alpha_{r,s,z}|^2 \qubit{k}\qubit{r}\qubit{s}\qubit{0}\qubit{0^{\ell}} + \qubit{k}\qubit{\Delta_{k,r}}.
\]

%%%
%%%
\item[(6)] Focusing on the last two registers, if they contain $\qubit{0}\qubit{0^{\ell}}$, then we multiply the content $s$ of the third register by $k$ to obtain $k\cdot s$ (seen as a scalar multiplication of a vector); otherwise, do nothing. Note that $k\cdot s$ is in $(\field_q)^m$ since $s\in(\field_{q})^m$ and $k\in\field_{q}$. Let $\qubit{\psi_6}$ denote the obtained quantum state.

%%%
%%%
\item[(7)]
Similarly, whenever $\qubit{0}\qubit{0^{\ell}}$ appears  in the last two registers, apply the inverse of the quantum Fourier transform  $(F_q^{-1})^m$ to the third register. This transform produces the quantum state
$
\qubit{\psi_7} =  \sum_{w\in(\field_q)^m} \gamma_{k,r,w} \qubit{k}\qubit{r}\qubit{w}\qubit{0}\qubit{0^{\ell}} + \qubit{k}\qubit{\Delta_{k,r}},
$
where $\gamma_{k,r,w}$ is a complex number given as
\[
\gamma_{k,r,w} =  \bra{r}\bra{w}\bra{0}\bra{0^{\ell}} (F \qubit{\psi_6}) = \frac{1}{q^{m}} \sum_{s\in(\field_q)^m}\sum_{z\in\field_q} \omega_q^{k(z - w\cdot s)}|\alpha_{r,s,z}|^2,
\]
where $F = I\otimes I\otimes (F_q^{-1})^m\otimes I$.

\item[(8)] Observe the last register in state $\qubit{0}\qubit{0^{\ell}}$ and discard the term $\qubit{\Delta_{k,r}}$. Finally, output the final quantum state
$
\qubit{\psi_8} = \sum_{w\in(\field_q)^m} \gamma_{k,r,w}  \qubit{k} \qubit{r}\qubit{w}.
$
This finishes the description of $\AAA$.
\end{enumerate}
\vspace*{1mm}
\hrule%\vspace*{5mm}
\vspace*{5mm}
%%%%%%%%%%%%%%%%%%%%%%%%%%%%%%%%%%%%

We define $\OO=\{O_k\}_{k\in[q-1]}$, where each $O_k$ is  a generalized quantumly corrupted codeword that
is realized by $\AAA$ with $\qubit{k}$ in the first register.

\sloppy To complete the proof, we need to estimate the average presence $av\Pre_{\OO}(C_x)$ of $C_x$ in $\OO$. For each index $k\in[q-1]$, the presence of $C_x$ in $O_k$ is exactly $\Pre_{O_k}(C_x) = q^{-m} \sum_{r\in\field_{q^m}}
|\gamma_{k,r,C_x(r)}|^2$, which equals $q^{-m}\sum_{r\in\field_q}\left| q^{-m} \sum_{s}\sum_{z}\omega_q^{k(z-w\cdot s)}|\alpha_{r,s,z}|^2 \right|^2$.  It thus follows that
\begin{eqnarray*}
av\Pre_{\OO}(C_x) &=& \frac{1}{q-1} \sum_{k\in[q-1]}\frac{1}{q^m}\sum_{r\in\field_{q^m}}
\left| \frac{1}{q^m} \sum_{z\in\field_q} \sum_{s\in(\field_q)^m}\omega_q^{k(z-D_x(r,s))}|\alpha_{r,s,D_x(r,s)}|^2\right|^2 \\
&=& \frac{1}{q^m(q-1)}  \sum_{k} \sum_{r}
 \left|  \sum_{j\in\field_q} \omega_q^{k\cdot j}
\left( \frac{1}{q^m}\sum_{s} |\alpha_{r,s,D_x(r,s)+j}|^2 \right)\right|^2 \\
&\geq&  \frac{1}{q^{2m}(q-1)^2}  \left| \sum_{k} \sum_{r}
\sum_{j\in\field_q} \omega_q^{k\cdot j}
\left( \frac{1}{q^m}\sum_{s} |\alpha_{r,s,D_x(r,s)+j}|^2 \right)\right|^2,
\end{eqnarray*}
where the last inequality follows from $\sum_{i=1}^{n}a_i^2\geq \frac{1}{n}\left(\sum_{i=1}^{n}a_i\right)^2$.
We therefore obtain
\begin{eqnarray*}
av\Pre_{\OO}(C_x) &\geq& \frac{1}{(q-1)^2}  \left|  \sum_{k\in[q-1]}
\sum_{j\in\field_q} \omega_q^{k\cdot j}
\left( \frac{1}{q^{2m}} \sum_{r\in\field_q} \sum_{s\in(\field_q)^m} |\alpha_{r,s,D_x(r,s)+j}|^2 \right)\right|^2.
\end{eqnarray*}
For each index $j\in\field_{q}$, we write $\beta_{j}$ for $q^{-2m}\sum_{r}\sum_{s}|\alpha_{r,s,D_x(r,s)+j}|^2$. Similarly to the proof of Lemma 4.5 of Kawachi and Yamakami \cite{KY06}, we can derive
\begin{eqnarray*}
av\Pre_{\OO}(C_x) &\geq& \frac{1}{(q-1)^2}  \left|  \sum_{k\in[q-1]}\beta_{0} + \sum_{1\leq j<q}\left(\sum_{k\in[q-1]}\omega_{q}^{k\cdot j}\right)\beta_j\right|^2 \\
&=& \frac{1}{(q-1)^2}  \left|  (q-1)\beta_{0} - \sum_{1\leq j<q}\beta_j\right|^2 \;\;=\;\; \frac{1}{(q-1)^2}  \left|  q\beta_0-1\right|^2,
\end{eqnarray*}
because $\sum_{j\in\field_q}\beta_j=1$ and $\sum_{k\in\field_{q}}\omega_q^{k\cdot j}= 0$ for any $j\neq0$. Since $\Pre_{O_D}(D_x) = q^{-2m}\sum_{r,s\in(\field_q)^m}|\alpha_{r,s,D_x(r,s)}|^2 =\beta_0$, it follows that
\[
av\Pre_{\OO}(C_x) \geq \frac{1}{(q-1)^2}  \left|  q\cdot \Pre_{O}(D_x) - 1  \right|^2  = \left( \frac{q}{q-1}\right)^2 \left( \Pre_{O_D}(D_x) - \frac{1}{q} \right)^2.
\]
This completes the proof of Lemma \ref{from-D-to-C}.
\end{proofof}

Next, we give the remaining proof of Lemma \ref{from-C-to-D}.

\begin{proofof}{Lemma \ref{from-C-to-D}}
Recall that $D = \mathrm{HAD}^{(q,m)}\odot C$ for a given $(q^m,n/m)_{q^m}$-code
$C$, provided that $n/m\in\nat^{+}$. Regarding this code $C$, a quantumly corrupted codeword $O_{C}$ is assumed to  act as
$O_C\qubit{r}\qubit{0}\qubit{0^d} = \sum_{z\in\field_{q^m}}\alpha_{r,z}\qubit{r}\qubit{z}\qubit{\phi_{r,z}}$.
Using this $O_C$ as an oracle, let us consider a polynomial-time quantum algorithm $\AAA$ defined below.   Let $e$ be the size of garbage qubits  produced in the description of $\AAA$.

%%%%%%%%%%%%%%%%%%%%%%%%%%%%%%%%%%%
\vspace*{5mm}
\hrule\vspace*{2mm}
\n{\sc Quantum Algorithm $\AAA$:}
\renewcommand{\labelitemi}{$\circ$}
\begin{enumerate}\vs{0}
  \setlength{\topsep}{-2mm}%
  \setlength{\itemsep}{0mm}%
  \setlength{\parskip}{0cm}%
%%%%%%%%%%%%%%%%%%%%%%%%%%%%%%%%%%%%

\item[(1)] Start with the quantum state $\qubit{\psi_1} = \qubit{r}\qubit{s}\qubit{0}\qubit{0^d}\qubit{0^e}$, where $r,s\in\field_{q^m}$.

\item[(2)] Change the register order to obtain the quantum state $\qubit{\psi_2} = \qubit{r}\qubit{0}\qubit{0^d}\qubit{s}\qubit{0^e}$.

\item[(3)] \sloppy Invoke $O_C$ using the first three registers and obtain the quantum state
$
\qubit{\psi_3} = \sum_{z\in\field_{q^m}}\alpha_{r,z}\qubit{r}\qubit{z}\qubit{\phi_{r,z}}\qubit{s}\qubit{0^e}.
$

\item[(4)] Compute the value $u= z\cdot s\;\mathrm{mod}\;q$ in a reversible fashion from $(s,z)$. We then obtain the quantum state
$
\qubit{\psi_4} = \sum_{z\in\field_{q^m}}\alpha_{r,z}\qubit{r}\qubit{z}\qubit{\phi_{r,z}}\qubit{s}\qubit{u}\qubit{\phi'_{s,z}},
$
where $\qubit{\phi'_{s,z}}$ indicates a certain garbage that might be  produced while reversing the computation for $u$ on a quantum computer.

\item[(5)] Again, change the register order so that we obtain the quantum state
$
\qubit{\psi_5} = \sum_{z\in\field_{q^m}}\alpha_{r,z}\qubit{r}\qubit{s}\qubit{z\cdot s\;\mathrm{mod}\;q} \qubit{z}\qubit{\phi_{r,z}}\qubit{\phi'_{s,z}}.
$
Finally, output $\qubit{\psi_5}$.
\end{enumerate}
\vspace*{1mm}
\hrule%\vspace*{5mm}
\vspace*{5mm}
%%%%%%%%%%%%%%%%%%%%%%%%%%%%%%%%%%%%

The desired quantumly corrupted codeword $O$ for $D$ is defined as
\begin{eqnarray*}%\eject
O\qubit{r}\qubit{s}\qubit{0^l}\qubit{0^d}\qubit{0^e}
&=&
\sum_{z\in\field_{q^m}}\alpha_{r,z}\qubit{r}\qubit{s}\qubit{z\cdot s\;\mathrm{mod}\;q} \qubit{z}\qubit{\phi_{r,z}}\qubit{\phi'_{s,z}}  \\
&=& \sum_{w\in\field_q}\qubit{r}\qubit{s}\qubit{w} \otimes \left( \sum_{z\in A_{s}(w)} \alpha_{r,z} \qubit{z}\qubit{\hat{\phi}_{r,s,z}} \right),
\end{eqnarray*}
where  $\qubit{\hat{\phi}_{r,s,z}} = \qubit{\phi_{r,z}}\qubit{\phi'_{s,z}}$ and  $A_s(a) = \{z\in\field_{q^m} \mid z\cdot s \equiv a\;\mathrm{mod}\;q\}$ for any $a\in\field_{q}$.
It is obvious that $O$ can be realized by $\AAA$.

To end the proof, we want to show that $\Pre_{O}(D_x)$ equals  $1/q+ (1-1/q)\Pre_{O_C}(C_x)$. For convenience, let the notation
 $T_r$ for each index  $r\in\field_{q^m}$ express the value $\sum_{s\in\field_{q^m}}\|\sum_{z\in A_{s}(D_x(r,s))}  \alpha_{r,z}\qubit{z}\qubit{\hat{\phi}_{r,s,z}}\|^2$.
 With this notation, the presence $\Pre_{O}(D_x)$ can be expressed as $q^{-m}\sum_{r\in\field_{q^m}}T_{r}$, which
equals  $\sum_{s\in\field_{q^m}}\sum_{z\in A_{s}(D_x(r,s))}|\alpha_{r,z}|^2$.
Since the condition ``$z\cdot s\equiv D_x(r,s)\;\mathrm{mod}\;q$'' is equivalent to the condition 
``$z\cdot s\equiv C_x(r)\cdot s\;\mathrm{mod}\;q$,''
it follows that $T_r = \sum_{z\in\field_{q^m}}\sum_{s\in EQ_q(z,C_x(r))}|\alpha_{r,z}|^2$, where
$EQ_q(a,b) =\{s\in\field_{q^m} \mid a\cdot s \equiv b\cdot s \;\mathrm{mod}\;q\}$.
We therefore derive
\begin{eqnarray*}
T_r &=& |EQ_q(C_x(r),C_x(r))|\cdot|\alpha_{r,C_x(r)}|^2
 + \sum_{z: z\neq C_x(r)}|EQ_q(z,C_x(r))|\cdot|\alpha_{r,z}|^2 \\
&=& q^m|\alpha_{r,C_x(r)}|^2 + q^{m-1}\sum_{z:z\neq C_x(r)}|\alpha_{r,z}|^2 \\
&=&  q^m \left( \frac{1}{q} + \left( 1 - \frac{1}{q} \right)|\alpha_{r,C_x(r)}|^2 \right),
\end{eqnarray*}
where the second equality follows from the fact that $|EQ_q(a,b)|=q^{m-1}$ if $a\neq b$.
{}From the above relation, we obtain
\begin{eqnarray*}
\Pre_{O}(D_x) &=& \frac{1}{q^m}\sum_{r\in \field_{q^m}}T_r
\;\;=\;\; \frac{1}{q} + \frac{1}{q^m}\left( 1 - \frac{1}{q} \right) \sum_{r\in\field_{q^m}} |\alpha_{r,C_x(r)}|^2 \\
&=& \frac{1}{q} + \left( 1 - \frac{1}{q} \right) \Pre_{O_C}(C_x).
\end{eqnarray*}
This completes the proof of Lemma \ref{from-C-to-D}.
\end{proofof}

In the end, we have finished the proof of Proposition \ref{concatenated_code}.

%%%%%%%%%%%%%%%%%%%%%%%%%%%%%%%%%%
\section{Complexity of Generalized Reed-Solomon Codes}\label{sec:Reed-Solomon}

We have shown in Proposition \ref{concatenated_code} that the concatenated code family $C^{GRS\mbox{-}H}$ has an efficient quantum list decoder if the generalized Reed-Solomon (GRS) codes are efficiently quantum list decodable.
In order to verify Theorem \ref{high-rate}, however, it remains to claim that the
GRS-codes are efficiently quantum list decodable when the bias is relatively large. This claim will be proven as Lemma \ref{Reed-Solomon} in Section \ref{sec:simple-approach} in a more general fashion.
For a much smaller bias, in contrast, there seems little hope in finding an efficient quantum list decoder, based on the common belief that $\np$-complete problems have no efficient quantum algorithms. In Sections \ref{sec:NPIP}--\ref{sec:BDVP}, we will further show that the GRS-codes have natural connections to the {\em noisy polynomial interpolation problem} (NPIP) of Naor and Pinkas \cite{NP99} and a lattice problem, which we call the {\em bounded distance vector problem} (BDVP).

%%%%%%
\subsection{Polynomial Reconstruction}\label{sec:simple-approach}

Proposition \ref{concatenated_code} requires the existence of efficient quantum list decodability of a family of GRS-codes.
This assumption can be removed for an appropriate choice of parameters. Now, we  claim, in the following technical lemma, that the family of GRS-codes is indeed quantumly list decodable.

\begin{lemma}\label{Reed-Solomon}
For any number $n\in\nat$, assume that a prime number $q$ and  real numbers $\varepsilon,\delta\in(0,1)$ satisfy the following conditions: $2\leq n\leq q$  and
$
\varepsilon' + \left(1-1/q-\varepsilon'\right) \sqrt{\frac{\Delta}{1+q\varepsilon'}} <\varepsilon \leq 1- 1/q
$
for a certain number $\varepsilon'\in(0,1)$, where $\Delta = 2(n-1)\log(q^2/(1-\delta))$. There exists a quantum list decoder for a $(q,n,q-n+1)_q$-generalized Reed-Solomon code with bias $\varepsilon$ and confidence $\delta$ running in time polynomial in $(n,q,1/\delta,1/(1-\delta))$.
\end{lemma}

Combining Proposition \ref{concatenated_code} together with Lemma \ref{Reed-Solomon}, Theorem \ref{high-rate} follows immediately.  Before proving Lemma \ref{Reed-Solomon}, we briefly present the proof of Theorem \ref{high-rate}, which leads to Theorem \ref{c-q-list-decoding}.

\begin{proofof}{Theorem \ref{high-rate}}
Consider the concatenated code $C^{GRS\mbox{-}H}[n,q,\theta] = \mathrm{HAD}^{(q,m)}\odot \mathrm{GRS}^{(M,M\theta,M)}$ with parameters $n,q,\theta$ as specified in the theorem, where $M=q^m$ and $n=mq^m\theta$.
The premise of the theorem implies
\[
\left(1-\frac{1}{q}\right)^2\left[ \frac{1}{M} + \varepsilon'' + \left(1-\frac{1}{M}-\varepsilon''\right) \sqrt{\frac{\Delta}{1+M\varepsilon''}} \right] < \frac{\varepsilon^2}{2} + \frac{\varepsilon^2}{2} = \varepsilon^2,
\]
which further implies
$\varepsilon'' + \left(1-1/M-\varepsilon''\right) \sqrt{\Delta/ (1+M\varepsilon'')} < \frac{q^2\varepsilon^2}{(q-1)^2} - \frac{1}{M}$. Now, choose an appropriate real number $\varepsilon'\in(0,1)$ so that
(1) $\varepsilon'' + \left(1-1/M-\varepsilon''\right) \sqrt{\Delta/(1+M\varepsilon'')}  < \varepsilon'$ and
(2) $\varepsilon' \leq  \frac{q^2\varepsilon^2}{(q-1)^2} - \frac{1}{M}$ (or equivalently, $\varepsilon^2\geq \left(1-1/q\right)^2\left( 1/M +\varepsilon'\right)$).

{}From (1), Lemma \ref{Reed-Solomon} guarantees the existence of  a quantum list decoder $\AAA$ for $\mathrm{GRS}^{(M,M\theta,M)}$ with bias $\varepsilon'$ and confidence $\delta$ running in time polynomial in $(n,M,1/\delta,1/(1-\delta))$. With this quantum list decoder together with (2), Proposition \ref{concatenated_code} provides us with the desired quantum list decoder for $C^{GRS\mbox{-}H}$ with bias $\varepsilon$ and confidence $\delta$.
\end{proofof}

To complete the proof of Theorem \ref{high-rate}, what still remains to deal with is the proof of Lemma \ref{Reed-Solomon}. A direct use of a polynomial reconstruction algorithm of Guruswami and Sudan \cite{GS99} works well to prove this lemma. We will apply this classical algorithm after collecting enough information on possible values of a target ``polynomial'' by a simple application of random sampling, that is, performing measurement on all oracle answers.

\begin{proofof}{Lemma \ref{Reed-Solomon}}
Let $n$ be an arbitrary message length and choose four parameters  $q\in\nat^{+}$ and $\varepsilon,\varepsilon',\delta\in(0,1)$ that satisfy the premise of the lemma. For simplicity, we write $C$ for $\mathrm{GRS}^{(q,n,q)}$.
Let $O$ be any quantumly corrupted codeword for $C_x$, having the form $O\qubit{r}\qubit{s}\qubit{0^{\ell}} = \sum_{z\in\field_{q}} \alpha_{r,z}\qubit{r}\qubit{s\oplus z}\qubit{\phi_{r,z}}$ for certain complex numbers $\alpha_{r,z}$ and certain unit-norm quantum states $\qubit{\phi_{r,z}}$. Recall that the presence of $C_x$ in $O$ is $(1/q)\sum_{r\in\field_q}|\alpha_{r,C_x(r)}|^2$.
Here, we want to find all messages $x$ satisfying the inequality  $\Pre_{O}(C_x) \geq 1/q+\varepsilon$.

Fix a message $x$ arbitrarily and omit script ``$x$'' in the following argument.
Let us define two sets $A_{\varepsilon'}=\{r\in \field_q\mid |\alpha_{r,C_x(r)}|^2\geq 1/q+\varepsilon'\}$ and
$D_{\varepsilon'}=\{(r,y)\in\field_q^2\mid |\alpha_{r,y}|^2\geq 1/q+\varepsilon'\}$. Note that $C_x$ passes at least $|A_{\varepsilon'}|$ points in $D_{\varepsilon'}$.
First, we note that $|D_{\varepsilon'}|\leq q^2/(1+q\varepsilon')$. This upper bound is easily obtained from
\[
q^2 \geq \sum_{r}\sum_{y}|\alpha_{r,y}|^2 \geq \sum_{(r,y)\in D_{\varepsilon'}}|\alpha_{r,y}|^2 \geq |D_{\varepsilon'}|\left(\frac{1}{q}+\varepsilon'\right).
\]
The assumption
$\Pre_{O}(C_x)\geq 1/q+\varepsilon$ implies
\begin{eqnarray*}
\frac{1}{q}+\varepsilon &\leq& \Pre_{O}(C_x)
\;=\;  \frac{1}{q}\sum_{r\in A_{\varepsilon'}}|\alpha_{r,C_x(r)}|^2 + \frac{1}{q}\sum_{r\in \field_q-A_{\varepsilon'}}|\alpha_{r,C_x(r)}|^2 \\
&\leq& \frac{|A_{\varepsilon'}|}{q}  + \frac{q - |A_{\varepsilon'}|}{q}\left(\frac{1}{q}+\varepsilon' \right).
\end{eqnarray*}
This concludes that $|A_{\varepsilon'}|\geq (1-\gamma_{\varepsilon,\varepsilon'})q$, where $\gamma_{\varepsilon,\varepsilon'} = \frac{1-1/q-\varepsilon}{1-1/q-\varepsilon'}$.

For a later use, we set $T' = \frac{q^2\log(q^2/(1-\delta))}{1+q\varepsilon'}$. Now, we claim that $(1-\gamma_{\varepsilon,\varepsilon'})^2q^2 > 2(n-1)T'$.
This inequality is equivalent to $(\varepsilon-\varepsilon')^2 >  (1-1/q-\varepsilon')^2\frac{\Delta}{1+q\varepsilon'}$,
which directly follows from our assumption that $\varepsilon > \varepsilon' +  (1-1/q-\varepsilon')\sqrt{\Delta/(1+q\varepsilon')}$,  where $\Delta = 2(n-1)\log(q^2/(1-\delta))$.
Let us consider the following quantum algorithm.

Initially, from the quantum state $\qubit{0}\qubit{0}\qubit{0}$, we generate $\qubit{\psi_0} = (1/\sqrt{q})\sum_{r\in\field_q}\qubit{r}\qubit{0}\qubit{0}$.  By making a query to oracle $O$, we generate $\qubit{\psi_1} = (1/\sqrt{q})\sum_{r}\sum_{y}\alpha_{r,y}\qubit{r}\qubit{y}\qubit{\phi_{y}}$. Next, we measure the first two registers and obtain $(r,y)$ with probability $|\alpha_{r,y}|^2/q$.
Let us repeat these steps exactly $T$ times, where $T$ is the minimal positive integer satisfying $2T'\geq T\geq T'$. Since $T\geq T'$, we obtain
\begin{equation}\label{eqn:T}
T \geq \frac{q^2\log(q^2/(1-\delta))}{1+q\varepsilon'} \geq  \frac{ \log\left( 1+q\varepsilon' \right)\left( 1- \delta \right)/q^2}{\log\left( 1-1/q^2 -\varepsilon'/q\right)},
\end{equation}
where we use inequalities: $\log(1-z)<-z$ and $1+q\varepsilon'\geq 1$.
After receiving each  answer from $O$, we perform a measurement in the computational basis over $\field_q\times \field_q$ and store a point $(r,y)$ that is a result of this measurement.

Let $S_{\varepsilon'}$ indicate the set of all the obtained points. Clearly, $|S_{\varepsilon'}|\leq T$ holds.
Note that,  with probability $(1-|\alpha_{r,y}|^2/q)^{T}$,  each point $(r,y)$ is never observed during the procedure. Hence, the probability $P$ of obtaining all $(r,y)$'s in $D_{\varepsilon'}$ is lower-bounded by
\[
 P \geq 1 - \sum_{(r,y)\in D_{\varepsilon'}}\left( 1- \frac{|\alpha_{r,y}|^2}{q}\right)^T
\geq 1 - \frac{q^2}{1+q\varepsilon'}\cdot \left(1-\frac{1}{q^2}-\frac{\varepsilon'}{q}\right)^T \geq \delta,
\]
where  the last inequality follows from Eq.(\ref{eqn:T}).
Therefore, the probability that $S_{\varepsilon'}$ includes  $D_{\varepsilon'}$ is at least $\delta$.

Lastly, we wish to find all univariate polynomials $p$ of degree at most $n-1$ that lie on at least $|A_{\varepsilon'}|$ points in $S_{\varepsilon'}$. For this purpose, we run the well-known Guruswami-Sudan polynomial reconstruction algorithm.
Earlier, Guruswami and Sudan \cite{GS99} described  a deterministic algorithm $\AAA$ that solves in time polynomial in $(m,\log{q})$ the following {\em polynomial reconstruction problem}.

\ms
\n{\sc Polynomial Reconstruction Problem}
\renewcommand{\labelitemi}{$\circ$}
\renewcommand{\labelitemi}{$\circ$}
\begin{itemize}\vs{-2}
  \setlength{\topsep}{-2mm}%
  \setlength{\itemsep}{0mm}%
  \setlength{\parskip}{0cm}%

\item {\sc Input:} three positive integers $m',n',t$ and $m'$ points $\{(x_i,y_i)\}_{i\in[m']}\subseteq \field_q\times\field_q$.

\item {\sc Output:} all univariate polynomials $p$ of degree at most $n'$ that lie on at least $t$ points, provided that $t>\sqrt{m'n'}$.
\end{itemize}

To apply the  Guruswami-Sudan algorithm to our case, letting $n'=n-1$, $m'=|S_{\varepsilon'}|$, and $t=|A_{\varepsilon'}|$, we should demand
the requirement that  $|A_{\varepsilon'}|> \sqrt{(n-1)|S_{\varepsilon'}|}$. This requirement is met because the choice of our parameters $\varepsilon$ and $\varepsilon'$ implies that
\[
|A_{\varepsilon'}|\geq (1-\gamma_{\varepsilon,\varepsilon'})q > \sqrt{2(n-1)T'} > \sqrt{(n-1)T} \geq  \sqrt{(n-1)|S_{\varepsilon'}|}.
\]
Therefore, the algorithm $\AAA$ correctly produces a list that includes all the polynomials $p$ of degree at most $n-1$ satisfying  $|\alpha_{r,p(r)}|^2\geq 1/q+\varepsilon'$ for at least $|A_{\varepsilon'}|$ indices $r$. Concerning the efficiency of the algorithm, we note that the running time of $\AAA$ is bounded by a polynomial in $(q,n)$. As a consequence, the list produced by $\AAA$ includes all messages $x$ for which  $\Pre_{O}(C_x) \geq 1/q+\varepsilon$.

Since $\AAA$ is deterministic, we can execute it quantumly as well. In the end,  we produce the desired list with probability at least $\delta$ in time polynomial in $(n,q,1/\delta,1/(1-\delta))$.
\end{proofof}

Due to the random sampling of quantum states necessary to apply for the Guruswami-Sudan  algorithm, the total number of oracle queries made by the quantum algorithm described in the above proof of Lemma \ref{Reed-Solomon} is at most $T$, guaranteeing the confidence $\delta$.
An important open question is whether the same confidence $\delta$ can be achieved with a significantly \emph{fewer} (\eg  a constant number of) queries.

To apply the Guruswami-Sudan algorithm,  we have required the bias $\varepsilon$ in the proof of Lemma
\ref{Reed-Solomon} to be relatively large. One may wonder whether, even if the bias is relatively small,
there is another way to list-decode  the generalized Reed-Solomon codes from a quantumly corrupted codeword.
In the next proposition, we will show  that any {\em efficient} quantum list decoder for the generalized Reed-Solomon codes with small bias and high confidence can be used to solve all $\np$-problems {\em efficiently} on a quantum computer with high success probability, leading to $\np\subseteq\bqp$.

\begin{proposition}\label{GRS-imply-BQP}
Let $t(n)$ be any function from $\nat$ to $\nat$ with $t(n)\geq n$ for all $n\in\nat$. If, for any arbitrary bias $\varepsilon(n)$,  there exists a quantum list decoder $\AAA$ for the generalized Reed-Solomon codes with bias $\varepsilon(n)$ and confidence $2/3$ running in $t(n)$ time, then every NP-problem can be solved  in $n^{O(1)}t(n)$ time by a certain quantum algorithm with success probability at least $2/3$. In particular, if $\AAA$ runs in polynomial time, then $\np\subseteq\bqp$ holds.
\end{proposition}

\begin{proof}
We want to give a polynomial-time reduction from a certain suitable $\np$-complete problem to an $\varepsilon$-QLDP for the GRS-code with respect to a specific quantumly corrupted codeword, where $\varepsilon$ will be defined later. As a target  $\np$-complete problem, we choose the following restricted form of the {\em interpolation problem} discussed by Goldreich, Rubinfeld, and Sudan \cite{GRS95}.

\ms
\n{\sc Constrained Interpolation Problem (CIP)}
\renewcommand{\labelitemi}{$\circ$}
\begin{itemize}\vs{-2}
  \setlength{\topsep}{-2mm}%
  \setlength{\itemsep}{0mm}%
  \setlength{\parskip}{0cm}%

\item {\sc Input:} three numbers $d,e,m\in\nat^{+}$, a prime number $q$, and a set $A=\{(x_1,y_1),\ldots,(x_m,y_m)\}\subseteq \field_q\times\field_q$ of $m$ points, expressed appropriately in binary.

\item {\sc Requirement:} $d_A(x_i)=2$ for any index $i\in[m]$, where $d_A(x)=|\{y\mid (x,y)\in A\}|$.

\item {\sc Question:} is there any univariate polynomial $p$ over $\field_{q}$ of degree at most $d$ such that $p(x_i)=y_i$ for at least $e$ different $i$'s?
\end{itemize}

Note that, when $e=1$, we always take a polynomial $p$ satisfying $p(x_1)=y_1$. Therefore, in what follows, we  assume that $e\geq2$.

The problem CIP is clearly in $\np$ and it can be proven to be NP-hard.\footnote{This fact is observed by examining the reduction constructed by Goldreich, Rubinfeld, and Sudan \cite{GRS95} from the {\em subset sum problem}, which is known to be $\np$-complete.}
As a starting point, let $d,e,m\in\nat^{+}$, let $q$ be a prime number, and let $A=\{(x_1,y_1),\ldots,(x_m,y_m)\}\subseteq \field_q\times\field_q$ as an input to the CIP.
Let $\ell=(m-1)/2$.
Any polynomial $p(r) = \sum_{i=1}^{d+1}z_ir^{i-1}$ for any $r$ can be viewed as a codeword $\mathrm{GRS}^{(\ell,d+1,q)}_{z}$, where $z=z_1z_2\cdots z_{d+1}$. For convenience, since $d_{A}(x_i)=2$ for all $i$'s, the set $D= \{x_1,\ldots,x_m\}$ of {\em code locators} has cardinality exactly $\ell$. Without loss of generality, we assume that $\ell+1\leq q$.

Based on the set $A$, we wish to construct a
quantumly corrupted codeword $O$.
For  any point $(x,y)$ in $\field_q\times\field_q$, if $(x,y)\in A$, let $\alpha_{x,y}=1/\sqrt{d_A(x)}$; otherwise, let $\alpha_{x,y}=0$.
The amplitude set $\{\alpha_{x,y}\}_{x,y\in\field_{q}}$ defines $O$ as
$O\qubit{x}\qubit{s}\qubit{t} = \sum_{y\in \field_{q}}\alpha_{x,y}\qubit{x}\qubit{y\oplus s}\qubit{t}$. Define $\varepsilon = 1/q - e/2\ell$.

It is not difficult to show that, for any polynomial $p$ of degree $d$,
$p$ passes on at least $e$ points in $A$ if and only if the presence of  $p$ (seen as a codeword) in $O$ satisfies the inequalities:
\[
\Pre_{O}(\mathrm{GRS}^{(\ell,d+1,q)}_{z}) = \frac{1}{|D|}\sum_{x\in D}|\alpha_{x,p(x)}|^2 \geq
\frac{1}{\ell}\sum_{i=1}^{e}\frac{1}{2} = \frac{e}{2\ell}
= \frac{1}{q} + \varepsilon,
\]
provided that $p$ is identical with $\mathrm{GRS}^{(\ell,d+1,q)}_z$.  Therefore, solving the CIP can be reduced to solving the $\varepsilon$-QLDP for the GRS-code $\{\mathrm{GRS}^{(\ell,d+1,q)}\}_{\ell,d,q}$  with respect to $O$. Moreover, it takes only quantum polynomial-time to {\em realize} $O$ from the set $A$ (which is given as an input). Applying a $t(n)$-time quantum list decoder for the $\varepsilon$-QLDP with confidence $2/3$, we  can obtain a valid list of polynomials $p$. Obviously, the size of the obtained list is at most $t(n)$. Since the list may contain certain illegitimate polynomials, we need to check that every candidate $p$ passes on at least $e$ different points in $A$. If the list contains a legitimate polynomial, we output ``YES''; otherwise, output ``NO.''
This quantum algorithm solves the CIP with success probability at least $2/3$.

If this quantum algorithm runs in polynomial time, we can solve efficiently the CIP with high probability, leading to the inclusion $\np\subseteq \bqp$ because the CIP is $\np$-complete.
\end{proof}

Despite the power of quantum computation, it seems unlikely that
polynomial-time quantum algorithms can solve all the $\np$-problems with  high success probability. Proposition \ref{GRS-imply-BQP} thus leaves little hope for finding a ``polynomial-time'' quantum list decoder for the GRS-codes with a {\em  smaller bias}.
However, it seems a challenging task to determine the exact threshold of such a bias for efficient quantum list decoders to exist.

%%%%%%
\subsection{Noisy Polynomial Interpolation Problem}\label{sec:NPIP}

As Proposition \ref{GRS-imply-BQP} indicates, for the generalized Reed-Solomon (GRS) codes, we may not be able to obtain  a polynomial-time quantum list decoder having extremely small bias; however, it is still meaningful to study, for example, {\em subexponential-time} quantum list decoders with relatively small bias for the GRS-codes and thus to seek their applications  to  the field of computational cryptography. Here, we wish to propose one of those possible applications.

Earlier,
Naor and Pinkas \cite{NP99} studied the {\em noisy polynomial interpolation problem} (NPIP) as an intractable assumption for a new cryptographic primitive, called {\em oblivious polynomial evaluation}. We restate their noisy interpolation problem as a promise problem of finding a unique polynomial passing through exactly one point from each given set.

\ms
\n{\sc Noisy Polynomial Interpolation Problem (NPIP)}
\renewcommand{\labelitemi}{$\circ$}
\begin{itemize}\vs{-2}
  \setlength{\topsep}{-2mm}%
  \setlength{\itemsep}{0mm}%
  \setlength{\parskip}{0cm}%

\item {\sc Input:} three numbers $k,m,n\in\nat^{+}$, a prime number $q$, $n$ distinct points $\{x_1,x_2,\ldots,x_n\}$ in $\field_q$, and $n$ sets $S_1,\ldots,S_n$, each of which consists of exactly $m$ elements from  $\field_q$, where $k+1\leq n\leq q$.

\item {\sc Promise:} there exists a {\em unique} polynomial $p$ of degree at most $k$ such that, for each index $i\in[n]$, there exists exactly one  element $y\in S_i$ satisfying $p(x_i)=y$.

\item {\sc Output:} the hidden polynomial $p$.
\end{itemize}

Disappointingly, no polynomial-time algorithm has been so far known to solve this promise problem NPIP. Apparent similarity exists between this problem and the GRS-codes (see, \eg Roth \cite{Rot06}) and, in the following proposition, this similarity helps us solve the NPIP using suitable quantum list decoders for the GRS-codes if such list decoders are actually built.

\begin{proposition}
If, for any bias parameter $\varepsilon(n)$,  there exists a quantum list-decoder for any $\mathrm{GRS}^{(n,k+1,q)}$-code with bias $\varepsilon(n)$ and confidence $2/3$, then there exists a quantum algorithm that solves the NPIP with probability at least $2/3$.
\end{proposition}

\begin{proof}
Take $n$ distinct elements $X=\{x_1,\ldots,x_n\}\subseteq\field_{q}$ and $n$ sets $S_1,\ldots,S_n$ of $m$ elements each. Let us assume that the promise of the NPIP holds for a unique polynomial, say, $p^*$ of degree at most $k$. Note that $k,m,n\leq q$. We set the bias parameter $\varepsilon$ to be $1/m-1/q$, and let $S$ be $\bigcup_{i\in[n]}S_i$.

Here, we define the $\varepsilon$-QLDP for the $\mathrm{GRS}^{(n,k+1,q)}$-code  with respect to a quantumly corrupted codeword $O$, which is defined by
$
O\qubit{x_i}\qubit{0} = \frac{1}{\sqrt{m}}\sum_{y\in S} \alpha_{x_i,y}\qubit{x_i}\qubit{y}
$
for each index $i\in[n]$, where $\alpha_{x_i,y} = 1$ if $y\in S_i$ and $0$ otherwise.
We first claim that the unique polynomial $p^*$ satisfies the condition $\Pre_{O}(p^*)\geq 1/q+\varepsilon$. Since $|X|=n$, it follows that
\[
\Pre_{O}(p^*) = \frac{1}{|X|}\sum_{x\in X}|\alpha_{x,p^*(x)}|^2 = \frac{1}{n}\sum_{x\in X}\frac{1}{m}
= \frac{1}{q} + \left( \frac{1}{m} - \frac{1}{q}\right) \geq \frac{1}{q} +\varepsilon.
\]
Hence, $p^*$ has codeword presence at least $1/q+\varepsilon$.

The assumption of the proposition guarantees the existence of a quantum list decoder $\AAA$ that solves the $\varepsilon$-QLDP
with confidence $2/3$.
To {\em realize} $O$ from the given inputs $(x_1,\ldots,x_n,S_1,\ldots,S_n)$ of the NPIP, we generate the quantum state $O\qubit{x_i}\qubit{s}$ by   choosing $y$ in $S$ uniformly at random and then generating the amplitude $\alpha_{x_i,y}/\sqrt{m}$.
For the NPIP, let us consider the following quantum algorithm.
\begin{quote}\vs{-1}
Taking $(k,m,n,q)$, $(x_1,\ldots,x_n)$ and $S_1,\ldots,S_n$ as input instance,  run the quantum list decoder $\AAA$ using $O$ as an oracle. We then obtain a list of polynomials $p$ that satisfy $\Pre_{O}(p)\geq 1/q+\varepsilon$. Since the hidden polynomial $p^*$ must be in the list, we {\em deterministically} check, through this list, whether each polynomial passes exactly one point from each set $S_i$. The uniqueness of $p^*$ ensures that this algorithm eventually finds $p^*$.
\end{quote}\vs{-1}
It is not difficult to show that the above quantum algorithm solves
 the NPIP with success probability at least $2/3$ because $\AAA$ has confidence $2/3$.
\end{proof}

%%%%%%%%%%%%%%%%%%%%%%%%%%%%%%%%%%
\subsection{Bounded Distance Vector Problem}\label{sec:BDVP}

The previous section has sought out an application of a quantum list decoder for the generalized Reed-Solomon (GRS) codes. Here, we further intend to explore its relevant computational problems. Let us recall that codewords (viewed as functions) of the GRS codes can be identified with \emph{polynomials}. Since polynomials are closely related to certain types of lattice problems, by exploiting this relationship, we will introduce a specific lattice problem, which we preferably call the {\em bounded distance vector problem} (BDVP). Next, we will show that any quantum algorithm solving this BDVP with high probability yields, for any bias $\varepsilon$,  a quantum list decoder for  GRS-codes with  bias $\varepsilon$ and relatively high confidence. The problem BDVP is formally described as follows.

\ms
\n{\sc Bounded Distance Vector Problem (BDVP)}
\renewcommand{\labelitemi}{$\circ$}
\begin{itemize}\vs{-2}
  \setlength{\topsep}{-2mm}%
  \setlength{\itemsep}{0mm}%
  \setlength{\parskip}{0cm}%

\item {\sc Input:} a number $n\in\nat^{+}$, $m$ basis vectors $b_1,b_2,\ldots,b_m\in\integer^{n}$, and a radius $\xi\in \rational^{\geq0}$.

\item {\sc Implicit Input:} an oracle that, given a vector $v\in\integer^{n}$, returns the square of the weighted norm, $\|v\|^2=\sum_{j\in[n]}\lambda_j^2v_j^2$, where  $\lambda=(\lambda_j)_j\in[0,1]^n$ is a predetermined (but hidden) weight vector and $v=(v_1,\ldots,v_n)$.

\item {\sc Output:} a list that contains all vectors $v$ in the
lattice $L$ spanned by $\{b_1,b_2,\ldots,b_m\}$ for which $\|v\|^2\leq \xi$ holds.
\end{itemize}

In the next proposition, we show the aforementioned relationship  between the BDVP and quantum list decoding.

\begin{proposition}\label{BDVP-list-decoder}
If there exists a quantum algorithm that solves the BDVP with probability at least $2/3$, then, for any positive bias $\varepsilon$,  there exists a  quantum list decoder for the family of generalized Reed-Solomon codes with  bias $\varepsilon$  and confidence  $2/3$.
\end{proposition}

\begin{proof}
A basic idea of using {\em Lagrange's interpolation formulas}
in the following argument comes from Bleichenbacher and Nguyen \cite{BN00}. To prove the proposition, it suffices to construct a  quantum ``reduction''  to the BDVP from the $\varepsilon$-QLDP for the generalized Reed-Solomon code $\mathrm{GRS}^{(M,n,q)}$, where $\varepsilon$ is any positive bias.
Let us assume that the BDVP with a hidden weight vector is quantumly solvable with success probability at least $2/3$. We start with  an arbitrary input instance given to the $\varepsilon$-QLDP for $\mathrm{GRS}^{(M,n,q)}$.

Fix a message length $n$ arbitrarily. Let $\varepsilon$ be any positive bias and assume, without loss of generality, that $\varepsilon$ is a rational number.  Fix a set $D_M=\{x_1,x_2,\ldots,x_M\}$ of $M$ distinct code locators in $\field_q$ and express the Cartesian product $D_M\times \field_q$ as $\{(x_i,z_{j})\mid i\in[M],j\in[q]\}$.  Let $O$ denote any  quantumly corrupted codeword $O$ for $\mathrm{GRS}^{(M,n,q)}$ and assume that
$
O\qubit{x_i}\qubit{s}\qubit{0} = \sum_{j\in[q]}\alpha_{i,j}\qubit{x_i}\qubit{s\oplus z_j}\qubit{\phi_{i,j}},
$
where $\qubit{\phi_{i,j}}$ is a certain unit-norm quantum state.
Let $a=(a_1,a_2,\ldots,a_n)\in(\field_q)^n$ be any hidden message and let $p_{a}(x) = \sum_{k\in[n]}a_{k}x^{k-1}\;(\mathrm{mod}\;q)$ denote its codeword $\mathrm{GRS}^{(M,n,q)}_{a}$, which is a  polynomial over $\field_q$ of degree at most $n-1$. Now, assume that  $\Pre_{O}(p_a) = \frac{1}{M}\sum_{i\in[M]}|\alpha_{x_i,p_a(x_i)}|^2 \geq 1/q+\varepsilon$.

Next, we will define an instance to the BDVP. Firstly, we define our radius $\xi\in\rational$ as  $M(1-1/q-\varepsilon)$.
Secondly, we define a lattice $L$ spanned by certain basis vectors $\{b_1,b_2,\ldots,b_m\}$ as follows.
The {\em (special) Lagrange interpolation polynomials} corresponding to $D_M$ are $L_i(x) = \prod_{j\in[M]-\{i\}} \frac{x-x_j}{x_i-x_j}$ in $\field_q[x]$, which are polynomials of degree $M-1$, for each index $i\in[M]$. Every polynomial $L_i(x)$ satisfies the following property: $L_i(x_i)=1$ and $L_i(x_j)=0$ if $j\neq i$.  Here, we assume that $L_i(x)$ is of the form $\sum_{k\in[M]}c_{ik}x^{k-1}$ for certain constants $c_{ik}$ in $\field_q$.
Note that $p_a$ satisfies the {\em Lagrange's interpolation formula}:
\begin{eqnarray*}
p_{a}(x) &=& \sum_{i\in[M]} p_{a}(x_i)L_i(x)
\;=\; \sum_{i\in[M]}\sum_{j\in[q]} \delta_{ij}^{(a)}z_{j}L_i(x) \\
&=& \sum_{k\in[M]} \left( \sum_{i\in[M]}\sum_{j\in[q]} \delta^{(a)}_{ij}z_jc_{ik} \right) x^{k-1},
\end{eqnarray*}
where $\delta_{ij}^{(a)} = 1$ if $p_{a}(x_i) = z_{j}$ and $0$ otherwise. Obviously, for each fixed pair $i$ and $a$, it holds that  $\sum_{j\in[q]}\delta_{ij}^{(a)}=1$. The vector   $\delta^{(a)}=(\delta^{(a)}_{ij})_{ij}\in\integer^{qM}$ becomes our {\em target vector} in the desired lattice $L$ (which will be defined below).

We consider only vectors $d=(d_{ij})_{ij}\in\integer^{qM}$ satisfying the condition $deg\left(\sum_{k\in[M]}\left(\sum_{i\in[M]}\sum_{j\in[q]} d_{ij}z_jc_{ik}\right) x^{k-1}\right) \leq n$, which is equivalent to  $\sum_{i\in[M]}\sum_{j\in[q]} d_{ij}z_jc_{ik} = 0\;(\mathrm{mod}\;q)$ for every index $k\in[n+1,q]_{\integer}$. Moreover, $d$ should satisfy that   $\sum_{j=1}^{q}d_{ij}=\sum_{j=1}^{q}d_{i'j}$ for all pairs $(i,i')$. At last, the lattice $L$ is defined as the collection of all vectors $d=(d_{ij})_{ij} \in \integer^{qM}$ such that
\renewcommand{\labelitemi}{$\circ$}
\begin{enumerate}\vs{-1}
  \setlength{\topsep}{-2mm}%
  \setlength{\itemsep}{0mm}%
  \setlength{\parskip}{0cm}%

\item $\sum_{j\in[q]} d_{ij} =\sum_{j\in[q]}d_{i'j}\;(\mathrm{mod}\;q)$ for all pairs $i,i'\in[M]$; and

\item $\sum_{i\in[M]} \sum_{j\in[q]} d_{ij}z_jc_{ik} = 0\;(\mathrm{mod}\;q)$ for all $k\in[n+1,M]_{\integer}$.
\end{enumerate}\vs{-1}
It is not difficult to show that $L$ forms a lattice. It is important to note that the target vector $\delta^{(a)}$ belongs to $L$. {}From the definition of $L$, a suitable set of basis vectors $\{b_1,b_2,\ldots,b_m\}$ for $L$ can be found easily (see, \eg Bleichenbacher and Nguyen \cite{BN00}).

Finally, we introduce an oracle $O'$ for the BDVP.
To formulate this $O'$, it suffices to define its associated  weight vector  $\lambda=(\lambda_{ij})_{ij}\in[0,1]^{qM}$. For each point $(x_i,z_j)\in D_M\times \field_{q}$, let $\lambda_{i,j} = \sqrt{1- |\alpha_{x_i,z_j}|^2}$.
The {\em weighted norm} $\|d\|$ of a vector $d=(d_{ij})_{ij}\in L$ is thus calculated as
$
\|d\| = \sqrt{  \sum_{i,j}d_{ij}^2\lambda_{ij}^2} = \sqrt{  \sum_{i,j}d_{ij}^2 ( 1 - |\alpha_{x_i,z_{j}}|^2 )}.
$
Therefore, the square of the weighted norm of $\delta^{(a)}$ equals
\begin{eqnarray*}
\|\delta^{(a)}\|^2 &=& \sum_{i\in[M]}\sum_{j\in[q]}\left(\delta^{(a)}_{ij}\right)^2\left( 1- |\alpha_{x_i,p_a(x_i)}|^2 \right)
\;=\; M - \sum_{i\in[M]}|\alpha_{x_i,p_a(x_i)}|^2 \\
&=& M \left( 1 - \Pre_{O}(p_a) \right).
\end{eqnarray*}
Since $\xi=M(1-1/q-\varepsilon)$, it follows that $\|\delta^{(a)}\|^2\leq \xi$ iff $\Pre_{O}(p_a)\geq 1/q+\varepsilon$.

To solve the $\varepsilon$-QLDP for $\mathrm{GRS}^{(M,n,q)}$ with respect to $O$, we first compute the set of basis vectors $b_1,\ldots,b_m$ and the radius $\xi$ as defined above. We then solve the BDVP using the weight vector (given by the oracle $O'$) with success probability at least $2/3$. Let $v_1,\ldots,v_k$ be the resulted list of vectors in $L$. For each $v_i$, find $a_i\in(\field_q)^n$ such that $v_i = \delta^{(a_i)}$ by solving a set of linear equations. These $a_i$'s form a list that contains all messages satisfying $\Pre_{O}(p_a)\geq 1/q+\varepsilon$. Moreover, this list can be obtained  with probability at least $2/3$.

It is not difficult to show that the above-described  quantum algorithm indeed solves the $\varepsilon$-QLDP for $\mathrm{GRS}^{(M,n,q)}$. This completes the proof of Proposition \ref{BDVP-list-decoder}.
\end{proof}

%%%%%%%%%%%%%%%%%%%%%%%%%%%%%%%%%%%%%%%%%%
\section{An Application to Quantum Search Problems}\label{sec:search-problem}

Theorem \ref{c-q-list-decoding} has given an efficiently quantumly list-decodable code family $C$ over a fixed code alphabet that
has  polynomially small code rate; in addition, $C$ is also efficiently classically list decodable. This fulfills our primary goal of this paper. As the next goal, we will seek an application of such an interesting code family to computational complexity theory. Of all possible applications, we will choose an issue on  approximate solvability of  quantum search problems. For ease of description, we use the notation $\prob_{M}[M(x)=b]$ to denote the probability that observing the final configuration of a quantum algorithm $M$ starting with input $x$ results in $b$. Analogous to $\np$-search problems, an {\em NBQP-search problem} $\PP$ is formally defined as a triplet $(\Sigma^*,M,p)$, where
$M$ is a polynomial-time  quantum algorithm taking inputs from $\Sigma^*\times\Sigma^*$ and $p$ is a polynomial, together with the requirement that, for every $x\in\Sigma^*$ and every witness $y\in\Sigma^{p(|x|)}$, there exists a bit $b$ such that $\prob_{M}[M(x,y)= b]\geq 2/3$. For each $x\in\Sigma^*$, let $S_{x,M} = \{y\in\Sigma^{p(|x|)}\mid \prob_{M}[M(x,y)= 1]\geq 2/3\}$ be the set of {\em solutions} of $x$. For simplicity,  we fix our message alphabet $\Sigma$ to be $\{0,1\}$ throughout this section.

\ms
\n{\sc NBQP-Search Problem}
\renewcommand{\labelitemi}{$\circ$}
\begin{itemize}\vs{-2}
  \setlength{\topsep}{-2mm}%
  \setlength{\itemsep}{0mm}%
  \setlength{\parskip}{0cm}%

\item {\sc input:} a (binary) string $x$ of length $n$;

\item {\sc output:} a solution $y\in S_{x,M}$ for $x$ if $S_x\neq\setempty$. Otherwise, output $\bot$ (a special symbol not in $\Sigma$).
\end{itemize}\vs{-1}

Define $L_{M}=\{x\mid S_{x,M}\neq\setempty\}$. A {\em solution function} $f$ for the NBQP-search problem $\PP= (\Sigma^*,M,p)$ satisfies that (i) for every $x\in L_{M}$, $f(x)$ belongs to $S_{x,M}$ and (ii) for every $x\not\in L_{M}$, $f(x)=\bot$. We also introduce a class NBQP of decision problems as follows: a language $L$ belongs to NBQP if and only if
there exist a polynomial-time quantum algorithm $M$ and a polynomial $p$ for which $(\Sigma^*,M,p)$ is an NBQP-search problem and $L=L_M$.

We want to show that a certain NBQP-search problem cannot be solved even ``approximately'' if $\bqp\neq \mathrm{NBQP}$.

\begin{proposition}\label{NBQP-vs-BQP}
Assuming that $\bqp\neq \mathrm{NBQP}$, for every positive polynomial triplet $(p,p',p'')$ with $p'(n)>p(n)$ for all numbers $n\in\nat$, there exists an  NBQP-search problem $\PP=(\Sigma^*,M,p)$ that satisfies the following: for any solution function $f$ for $\PP$, no polynomial-time quantum algorithm $\BB$  finds strings $y$, on each input $x\in L_{M}$ of length $n$, with probability at least $1-\frac{2p(n)}{p'(n)(p(n)+2)}$ such that the relative distance $\Delta(y,f(x))$ is at most $1/2-1/p(n)$; on every input $x\not\in L_{M}$, $\BB$ outputs $\bot$ with probability at least $1/2+1/p''(n)$.
\end{proposition}

This proposition roughly implies that solving NBQP-search problems on average leads to solving them in worst case.
The proof of the proposition requires the following technical lemma, which gives a method of computing solution functions.
Recall from Section \ref{sec:input-output-model} the notation $(f(x))_i$.

\begin{lemma}\label{NBQP-search}
Let $s$ be any positive polynomial with $s(n)\geq 6$ for every $n\in\nat$. The following two statements are logically equivalent.
\renewcommand{\labelitemi}{$\circ$}
\begin{enumerate}\vs{-2}
  \setlength{\topsep}{-2mm}%
  \setlength{\itemsep}{0mm}%
  \setlength{\parskip}{0cm}%

\item For every $\mathrm{NBQP}$-search problem $\PP=(\Sigma^*,M,p)$, there exist its solution function $g$ and a polynomial-time quantum algorithm $\AAA$ such that (i) for every $x\in L_{M}$, $\prob_{\AAA,i}[\AAA(x,1^i)= (g(x))_i]\geq 1/2+ 1/s(|x|)$  and (ii)  for every $x\not\in L_{M}$,  $\prob_{\AAA,i}[\AAA(x,1^i)= 0]\geq 1/2+ 1/s(|x|)$, where ``$i$'' is a random variable uniformly distributed over $[p(n)]$.

\item For every $\mathrm{NBQP}$-search problem, there exist its solution function $f$ and a polynomial-time quantum algorithm $\BB$ such that, for every $x\in\Sigma^*$, $\prob_{\BB}[\BB(x)= f(x)]\geq 2/3$.
\end{enumerate}
\end{lemma}

With the help of the above lemma, we give the proof of  Proposition \ref{NBQP-vs-BQP}.

\begin{proofof}{Proposition \ref{NBQP-vs-BQP}}
We show the proposition by contradiction. First of all, we assume that $\bqp\neq \mathrm{NBQP}$.
Toward a contradiction, we assume that there exist a positive polynomial triplet $(p,p',p'')$ satisfying $p'(n)>p(n)$ for every $n\in\nat$ that meet  the following requirement: for any choice of  NBQP-search problem $\PP=(\Sigma^*,M,p)$, there are
a solution function $g$ for $\PP$  and a polynomial-time quantum algorithm $\BB$ for which (i)  on each input $x\in L_M$, $\BB$ finds with probability at least $1-\frac{2p(n)}{p'(n)(p(n)+2)}$  a string $y$ satisfying $\Delta(y,g(x))\leq 1/2-1/p(n)$ and  (ii) on every input $x\not\in L_M$, $\BB$ outputs $\bot$ with probability at least $1/2+1/p''(n)$.
Let us fix a polynomial $s$ satisfying that $s(n)\geq  \max\{6,p''(n),p'(n)p(n)/(p'(n)-p(n))\}$ for all numbers $n\in\nat$. Notice that $L_{M}$ belongs to NBQP.

We wish to compute $(g(x))_i$ from $(x,1^i)$ using $\BB$ so that we
obtain Lemma \ref{NBQP-search}(1). Let us consider the following algorithm $\AAA$: on input $(x,1^i)$, run the quantum algorithm $\BB$ on input $x$ and then output the $i$th bit of its outcome $y$ if $y\neq\bot$, and output $0$ otherwise.

Let $x$ be an arbitrary string of length $n$. If $x\in L_M$, then the average  probability of $\AAA$ producing $(g(x))_i$ correctly over all $i$'s is lower-bounded by
\begin{eqnarray*}
\prob_{\AAA,i}[\AAA(x,1^i)=(g(x))_i]
&\geq& \left(1-\frac{2p(n)}{p'(n)(p(n)+2)}\right)\left(1-\max_{y}\{\Delta(y,g(x))\}\right) \\
&\geq& \left(1-\frac{2p(n)}{p'(n)(p(n)+2)}\right)\left(1-\left(\frac{1}{2}-\frac{1}{p(n)}\right)\right) \\
&=& \frac{1}{2}+ \frac{1}{p(n)} - \frac{1}{p'(n)} \;\;\geq\;\; \frac{1}{2}+\frac{1}{s(n)},
\end{eqnarray*}
where the maximization is taken over all strings $y$ produced by $\BB$ that satisfy $\Delta(y,g(x))\leq 1/2-1/p(n)$.
If $x\not\in L_M$, then it follows that $\prob_{\AAA,i}[\AAA(x,1^i)=0] =  \prob_{\BB}[\BB(x)=\bot] \geq 1/2+1/p''(n) \geq 1/2+1/s(n)$.
Since $\PP$ is arbitrary, the statement of Lemma \ref{NBQP-search}(1) holds. Lemma \ref{NBQP-search}(2) then provides us with  a polynomial-time quantum algorithm that computes a certain solution function $f$  correctly with probability at least $2/3$. Since $L_{M} = \{x\mid f(x)\in\Sigma^*\}$ holds,  $L_{M}$ must be recognized with probability at least $2/3$ on a quantum computer in polynomial time; thus, $L_{M}$ belongs to $\bqp$. As a result, we conclude that  NBQP is included in BQP, a contradiction against our assumption that $\bqp\neq \mathrm{NBQP}$.
\end{proofof}

Finally, we present the proof of Lemma \ref{NBQP-search}, in which we extensively utilize an efficiently quantumly and classically list-decodable code family given in Theorem \ref{c-q-list-decoding}.

\begin{proofof}{Lemma \ref{NBQP-search}}
Let $s$ be any positive polynomial with $s(n)\geq6$ for every $n\in\nat$.
Because we consider only sufficiently large lengths $n$, we can assume without loss of generality that, for a certain fixed constant $k\geq1$, $s(n) = n^k$ holds for all numbers $n\geq6$. By Theorem \ref{c-q-list-decoding}, there are  a polynomial-time computable function $t$ and a $(t(n),n)_2$-code family $C$ that has a polynomial-time quantum list decoder $\DD$, with bias $1/s(n)$ and confidence $2/3$, producing a list of message candidates, where $n$ is a message length. Let $q$ denote a positive polynomial that bounds the sizes of any valid list produced by $\DD$. For convenience, we also assume that $t(n)\geq n$ for all numbers $n\in\nat$.
Moreover, we write $D$ for a polynomial-time classical list decoder for $C$. Note that, for each $y$, $C_y$ denotes the codeword, to which $y$ is encoded, of block length $t(|y|)$.
For the sake of convenience, in this proof, we also identify this codeword $C_y$ (defined as a function in Section \ref{sec:block-code}) as a $t(n)$-letter string $C_y(0)C_y(1)\cdots C_y(t(n)-1)$.

The implication (2) $\Rightarrow$ (1) in the lemma is trivial, since  $1/2+1/s(n)\leq 2/3$ and, if we can computer $f(x)$ with high probability, then we can compute its $i$th bit $(f(x))_{i}$ or the symbol $\bot$ with  success probability at least $2/3$. Hereafter, assuming (1), we intend to show (2). Let $\PP=(\Sigma^*,M,p)$ be any NBQP-search problem. To make our proof simple, we assume that $p(n)\geq n$ for all numbers $n\in\nat$. First, we reduce the error probability of the quantum algorithm $M$ to be exponentially small (without changing the witness size). This step can be done by a standard technique of {\em majority voting} among polynomially many runs of the original quantum algorithm. To be more precise, there exists a polynomial-time quantum algorithm $M'$, depending only on $(p,r,M)$, that satisfies the following two conditions:
\renewcommand{\labelitemi}{$\circ$}
\begin{enumerate}\vs{-1}
  \setlength{\topsep}{-2mm}%
  \setlength{\itemsep}{0mm}%
  \setlength{\parskip}{0cm}%

\item for every $x\in L_{M}$ and every $y\in S_{x,M}$,
$\prob_{M'}[M'(x, y) = 1] \geq 1- 2^{-r(|x|)}$; and

\item for any other pair $(x,y)$ with $y\in\Sigma^{p(|x|)}$,
 $\prob_{M'}[M'(x, y) = 0] \geq 1- 2^{-r(|x|)}$,
\end{enumerate}\vs{-1}
where $r(n)=q(p(n))+3$. Notice that $L_{M'}$ coincides with $L_{M}$.

Let us consider a quantum search problem $\PP'=(\Sigma^*,N,p)$ defined by
the following quantum algorithm $N$.
\begin{quote}
On input $(x,z)$ with $n=|x|$, if $|z|\neq t(p(n))$, then reject the input immediately. Otherwise, run the classical list decoder $D$ in polynomial time using $z$ as a classically corrupted codeword (or a received word) to produce with probability at least  $5/6$ a list $T$ of message candidates for $C$. Check deterministically whether $z=C_y$ holds for a certain string $y$ in $T$. If there is no such $y$, reject the input. On the contrary, if  $z=C_y$, then run $M'$ on the input $(x,y)$ and outputs its outcome.
\end{quote}

First, we claim that $\PP'$ is indeed an NBQP-search problem.
Fix an arbitrary $n\in\nat$, take any $x\in\Sigma^n$, and consider the case in which $x\in L_{M}$. Since there exists a witness $y\in\Sigma^{p(n)}$ for $x$, $y$ should be included in the list $T$. Hence,  its corresponding codeword $z=C_y$ forces $N$ to accept $(x,z)$ with probability at least $\frac{5}{6}\left(1-2^{-r(|x|)}\right)\geq 2/3$, because $r(n)\geq3$. For the other case where $x\not\in L_{M}$, let $z$ be any string in $\Sigma^{t(p(n))}$.  If $z\neq C_y$ for all $y\in T$, then $N$ rejects $(x,z)$ with probability at least $5/6$. By contrast, if $z=C_y$ holds for a certain $y\in T$, then $N$ accepts $(x,z)$ with probability $\leq \frac{5}{6}\cdot 2^{-r(|x|)}\leq 1/3$. Therefore, $\PP'$ is an NBQP-search problem.

Again, applying the majority vote technique, we can reduce the error probability of $N$ down to $2^{-r(n)}$. Abusing the notation, we use the same notation $N$ to denote this new algorithm.
For our NBQP-search problem $\PP'$, the statement  (1) gives a solution function $g$  and a polynomial-time quantum algorithm $\AAA$ for which
$\prob_{\AAA,i}[\AAA(x,1^i)=(g(x))_i] \geq 1/2+ 1/s(n)$ for every $x\in L_{M}\cap\Sigma^n$, and $\prob_{\AAA,i}[\AAA(x,1^i)=0]\geq 1/2+1/s(n)$ for every $x\in \Sigma^n- L_{M}$.
Now, assume that the final quantum state $\AAA\qubit{x,1^i}$ has the form
\[
\AAA\qubit{x,1^i} = \alpha_{x,i,0} \qubit{i}\qubit{0}\qubit{\phi_{x,0}} + \alpha_{x,i,1} \qubit{i}\qubit{1}\qubit{\phi_{x,1}}
\]
with certain amplitudes $\{\alpha_{x,i,b}\}_{x,i,b}$, where $\|\qubit{\phi_{x,b}}\|=1$ for any bit $b$. It is obvious that $\prob_{\AAA,i}[\AAA(x,1^i)=(g(x))_i] = (1/p(n))\sum_{i\in[p(n)]}|\alpha_{x,i,(g(x))_i}|^2$ for $x\in L_M$ and $\prob_{\AAA,i}[\AAA(x,1^i)=0] = (1/p(n))\sum_{i\in[p(n)]}|\alpha_{x,i,0}|^2$ for $x\not\in L_M$.

We fix an arbitrary $x\in L_{M}$ of length $n$ and, in the meantime, we omit script ``$x$.''
Let us define an oracle $O$ as
\[
O \qubit{i}\qubit{e}\qubit{0} = \alpha_{i,0}\qubit{i}\qubit{e\oplus 0}\qubit{\phi_{i,0}} + \alpha_{i,1}\qubit{i}\qubit{e\oplus 1}\qubit{\phi_{i,1}}
\]
for any $e\in\{0,1\}$ and any $i\in[p(n)]$.  This oracle $O$ is a quantumly corrupted codeword for $C$ and obviously $O$ can be realized by $\AAA$. If there exists a string $y$ satisfying $C_y=g(x)$, then the presence of $C_y$ in $O$ is calculated as
\[
\Pre_{O}(C_y) = \frac{1}{p(n)}\sum_{i}|\alpha_{i,C_y(i)}|^2
= \frac{1}{p(n)}\sum_{i}|\alpha_{i,(g(x))_i}|^2 \geq \frac{1}{2} + \frac{1}{s(n)}.
\]
This makes us possible to run $\DD$ using $O$ to list-decode $C$.

Finally, we define a new quantum algorithm $\BB$, based on the quantum list decoder $\DD$ for $C$, that finds a witness of the problem $\PP'$. Recall that $\DD$ produces a list of size at most $q(n')$ for each message length $n'$. We assume the standard lexicographic order in $\Sigma^{p(n)}$. Let us consider the following quantum algorithm $\BB$.
\begin{quote}
On input $x$ ($n=|x|$), run $\DD$ using $O$ as an oracle to produce a list $T'$ of at most $q(p(n))$ message candidates (since the message size is $p(n)$), which includes the solution $g(x)$ (if $x\in L_M$) or the string $0^{p(n)}$  (if $x\not\in L_M$), with probability at least $1-2^{-r(n)}$. Run $N$ on the input $(x,z)$ sequentially for all elements $z\in T'$ in order. Output the lexicographically smallest $z\in T'$ for which $N(x,z)$ outputs $1$ if any. On the contrary, if there is no such $z$, output $\bot$.
\end{quote}
Let $f(x)$ denote the minimal string $z$ in $T'$ such that (i) $\prob_{N}[N(x,z)=1]\geq 1-2^{-r(n)}$ and (ii)  $\prob_{N}[N(x,z')=0]\geq 1-2^{-r(n)}$ for all $z'<z$ in $T'$ if any; let $f(x)=\bot$ otherwise.  Since $|T'|\leq q(p(n))$, the probability that $\BB$ on input $x$ of length $n$ outputs $f(x)$ correctly is lower-bounded by
\[
\left(1-2^{-r(n)}\right)^{q(p(n))} \geq 1 - 2^{-r(n)+q(p(n))-1}
\geq 3/4.
\]
This guarantees that the  success probability of obtaining $f(x)$ is at least $3/4$. Since $\PP$ is arbitrary, the statement (2) should hold.
\end{proofof}

%%%%%%%%%%%%%%%%%%%%%%%%%%%%%%%%%
\section{Local Quantum List Decoding}\label{sec:local-list-decoding}

The previous sections have dealt with a specific computational model using implicit inputs and explicit outputs. When the running time of a quantum list decoder is limited to {\em sublinear}, however, it becomes impossible to produce a short list of messages {\em explicitly}. In such a case, it is better to allow the quantum list decoder to produces a list of short ``descriptions'' of  \emph{oracle quantum circuits},  each of which can generate every block symbol of a specific message by means of  an appropriate oracle access to a given quantumly corrupted codeword. We call such a model an {\em implicit-input implicit-output model}.
We will discuss a realm of quantum list decoding on this specific model and briefly state two results on the hardness amplification of quantum circuits.

Let us first introduce the notion of {\em local quantum list decoding}, analogous to the well-known notion of local list decoding.

\begin{definition}[local quantum list decoding]
Let $C$ be any $(M(n),n,d(n))_{q(n)}$-code family with a message alphabet $\Sigma$ (not depending on the choice of $n$). We say that $C$ is {\em locally quantum list decodable} with bias $\varepsilon$ and confidence $\delta$ if there exists a quantum algorithm $\AAA$ such that, for any message length $n\in\nat$, any quantumly corrupted codeword $O$ for $C$, and any message $x=x_1x_2\cdots x_n\in\Sigma^n$ satisfying  $\Pre_{O}(C_x)\geq 1/q+\varepsilon(n)$, the following two
conditions hold with probability at least $3/4$:
\renewcommand{\labelitemi}{$\circ$}
\begin{enumerate}\vs{-2}
  \setlength{\topsep}{-2mm}%
  \setlength{\itemsep}{0mm}%
  \setlength{\parskip}{0cm}%

\item $\AAA(n)$ outputs a list of ``descriptions'' of $\ell$
oracle quantum circuits  $D_1,D_2,\ldots,D_{\ell}$; and

\item there exists an index $j\in[\ell]$ such that, for every index $i\in[n]$ (expressed in binary),  $D_j^{O}$ on input $i$ outputs $x_i$ with probability at least $\delta(n)$
\end{enumerate}
\end{definition}

Similarly to the concatenated code family $C^{GRS\mbox{-}H}$, we can define another concatenated code family $C^{RM\mbox{-}H}$ using appropriate Reed-M{\"u}ller codes instead of the generalized Reed-Solomon codes. Following an argument of Sudan, Trevisan, and Vadhan \cite{STV01}, we can claim that the code $C^{RM\mbox{-}H}$ is efficiently locally quantumly list decodable with polynomially small bias and confidence $2/3$. For the proof of this claim, by  Lemma \ref{from-D-to-C},  it suffices for us to construct  an efficient quantum list decoder for the Reed-M{\"u}ller codes by following Sudan, Trevisan, and Vadhan \cite{STV01}. Such a quantum list decoder can be given by employing  an argument similar to that of Lemma \ref{Reed-Solomon}. Hence, we can conclude:

\begin{proposition}\label{local-decode}
There exists a code family of polynomially small rate and constant codeword alphabet size that are efficiently locally quantum list decodable with polynomially small bias and confidence $2/3$.
\end{proposition}

An immediate consequence of this proposition is the hardness amplification of quantum circuits, obtained by again following an argument of Sudan, Trevisan, and Vadhan \cite{STV01}.

\begin{corollary}
There exists a constant $d>0$ for which the following is true. Let $\varepsilon\in(0,1)$ and let $f$ be any Boolean function from $\{0,1\}^{k(n)}$ for a certain function $k(n)$. If no quantum circuit of size $s$ computes $f$ with success probability at least $\delta$, then there exists a Boolean function $g$ mapping $\{0,1\}^{\ell(k(n))}$ to $\{0,1\}$ with a certain function $\ell(n)\in n^{O(1)}$ such that no quantum circuit $C$ of size $s'=(k(n)/\varepsilon)^d\cdot s$ satisfies $\prob_{C,x}[C(x)=g(x)]\geq 1/2+\varepsilon$, where $C(x)$ denotes the random variable indicating the observed outcome bit of $C$ on input $x$.
\end{corollary}

%%%%%%%%%%%%%%%%%%%%%%%%%%%%%
\section{Concluding Remarks and Open Problems}

The main theme of this paper is to show the existence of a quantumly  list-decodable code family of polynomially small code rate over a fixed code  alphabet and to seek its application to computational complexity theory. To achieve such goals,  we  have considered certain codes made up of generalized Reed-Solomon (GRS) codes,  concatenated with the Hadamard codes, and we have proven that they are indeed efficiently quantum list decodable whenever the bias of their codeword presence is relatively large.
Notice that a core part of the proof of this result heavily relies on  a classical algorithm of Guruswami and Sudan \cite{GS99} and it therefore requires a relatively large number of queries. For certain types of applications, it may be desirable to make a fewer queries. At present, we have no answer to the question of whether there exists a quantum list decoder that makes a significantly fewer queries (say, less than the degree of a hidden polynomial).

Because of the different formulations of classical list decoding and quantum list decoding, we cannot verify that all classically list decodable codes are also quantumly list decodable. Among all codes of polynomially small rate, is there any quantum list decodable code that is not even classically list decodable?

When a bias becomes arbitrary small, in contrast, we have shown that the aforementioned concatenated code is unlikely to be efficiently quantumly list decodable, because
the GRS codes are unlikely to have efficient
quantum list decoders against arbitrary small bias. If we relax the running time of list decoders, can we build a  subexponential-time quantum list decoder for the GRS code against arbitrary bias? Another important open problem is to find useful applications of quantum list decoding to a wide range of topics in quantum information processing.

%%%%%%%%
%%%%%%%%
\section*{Appendix}

In this appendix, we will present the proofs of Propositions \ref{lower-bound-presence}--\ref{QU-upper-bound}, which have left unproven in Section \ref{sec:presence-distance}. The proof of Proposition \ref{lower-bound-presence} comes from an early result of Kawachi and Yamakami \cite{KY06} and the proof of Proposition \ref{QU-upper-bound} closely follows an argument of Guruswami, H{\aa}stad, Sudan, and Zuckerman \cite{GHSZ02}.

We begin with the proof of Proposition \ref{lower-bound-presence}.  Earlier, Kawachi and Yamakami \cite{KY06} presented a  relatively good upper bound on the size of a message list in terms of the value of codeword presence by employing a geometric method of Guruswami and Sudan \cite{GS00b}, who gave a $q$-ary extension of the well-known {\em Johnson bound}. Let $C$ be any $(M(n),n,d(n))_{q(n)}$-code family with a message space $\Sigma_n$ and define $P_{q(n)}(M(n),d(n),\varepsilon(n))$ as $\sup_{O}\left\{\left|\{x\in\Sigma_n\mid \Pre_{O}(C_x)\geq  \varepsilon(n)\}\right|\right\}$, where ``$\mathrm{sup}$'' is taken over all quantumly corrupted codeword $O$ for $C$. The following statement is a slight modification of Lemma 3.4 of Kawachi and  Yamakami \cite{KY06} and we therefore omit its proof.

\begin{lemma}\label{johnson-bound}
Let $n$ be any message length. Let $(\varepsilon(n),q(n),d(n),M(n))$ satisfy the inequality $\varepsilon(n) > \ell(n)$, where $\ell(n)$ equals $1/q(n) + \left(1 - 1/q(n) \right) \sqrt{1 -  (d(n)/M(n)) \left( q(n)/(q(n)-1)\right) }$. Assume that $C$ is an $(M(n), n, d(n))_{q(n)}$-code family.
The value $P_{q(n)}(M(n),d(n),\varepsilon(n))$ is upper-bounded by
$\min\left\{ M(n)(q(n) - 1), \frac{d(n) \left(1 - 1/q(n)\right)}{d(n) \left(1 - 1/q(n) \right) + M(n)\varrho(n)} \right\}$, where $\varrho(n) = (\varepsilon(n) - 1/q(n))^2 - \left(1 - 1/q(n)\right)^2$.
In the case of $\varepsilon(n) = \ell(n)$, it holds that
$P_{q(n)}(M(n),d(n),\varepsilon(n)) \leq 2M(n)(q(n) - 1) - 1$.
\end{lemma}

 Proposition \ref{lower-bound-presence} easily follows from Lemma \ref{johnson-bound}.

\begin{proofof}{Proposition \ref{lower-bound-presence}}
Consider any $(M(n),n,d(n))_{q(n)}$-code family $C$. Since $\lambda$ is the relative distance of $C$, it follows that $\lambda = d(n)/M(n)$. For readability, we will omit the parameter ``$n$'' in the following calculation. Let $c>0$ be a constant and suppose that the upper bound of $P_q(M,d,\varepsilon)$ given in Lemma \ref{johnson-bound} does not exceed $an^c$; that is,
assuming $\varepsilon>\ell$, it holds that
\[
P_{q}(M,d,\varepsilon) \leq \frac{d \left(1 - 1/q\right)}{d \left(1 - 1/q \right) + M\varrho} \leq an^c.
\]
{}From the last inequality, we immediately obtain
\[
M\left( \varepsilon - \frac{1}{q} \right)^2 \geq \frac{d \left(1 - 1/q\right)}{an^c} - d \left(1 - \frac{1}{q} \right) + M \left(1 - \frac{1}{q} \right)^2,
\]
since $\varrho = (\varepsilon-1/q)^2-(1-1/q)^2$. The absolute
value $|\varepsilon - 1/q|$ is thus lower-bounded by
\[
|\varepsilon - \frac{1}{q}| \geq \left( \frac{d \left(1 - 1/q\right)}{M an^c} - \frac{d\left(1 - 1/q \right)}{M} + \left(1 - \frac{1}{q} \right)^2 \right)^{1/2}.
\]
Assuming that $\varepsilon\geq 1/q$, we therefore conclude
\begin{eqnarray*}
\varepsilon &\geq& \frac{1}{q} + \left(1 - \frac{1}{q} \right)  \left( 1   - \frac{d}{M\left(1 - 1/q \right)} + \frac{d}{M an^c \left(1 - 1/q\right)} \right)^{1/2}.
\end{eqnarray*}
Moreover, in the case of $\varepsilon =\ell$, the definition of $\ell$ yields  $\varepsilon = \frac{1}{q} + \left(1 - \frac{1}{q} \right)  \left( 1   - \frac{d}{M\left(1 - 1/q \right)} \right)^{1/2}$.
Since $QL^{poly}_c(\lambda)\geq \varepsilon$, the proposition follows immediately from the relation $\lambda = d/M$.

The second part of the proposition can be directly obtained by making $c$ approach to the infinity.
\end{proofof}

Next, we give the proof of Proposition \ref{QU-upper-bound}.

%%%
%%%

\begin{proofof}{Proposition \ref{QU-upper-bound}}
Let $q$ be any odd prime number and fix $c\in\nat^{+}$ and $R\in(0,1)$ arbitrarily to satisfy $c>2(q-1)$. Let $n=\floors{MR}$ and set $I_n=[0,M-1]_{\integer}$. In this proof, we  consider only linear  $(M,n)_{q}$-codes. Recall from Section \ref{sec:presence-distance} the notations $V_n$, $W_n$, and $E(w,\varepsilon)$.
For brevity, let $\varepsilon= QU^{const}_{c}(R)$ and set $\bfzero = 0^{M}$ and $t=q-1$.

Since $V_n$ is composed of all vectors $v=(v_{r,z})_{r\in I_n,z\in\field_q}$ with  $v_{r,z}\in[0,t]_{\integer}$  and $\sum_{z\in\field_q}v_{r,z}=t$ for every index $r\in I_n$, it follows that  $|V_n|\geq q^{M}$. Note that,
for every $v\in V_n$, the vector $\hat{v}=(v_{r,z}/t)_{r,z}$ belongs to $W_n$.  Write $\hat{v}_{r,z}$ for $v_{r,z}/t$.
Let the notation $V_{r,n}$ denote the $r$th block of $V_n$. Note that $V_{r,n}$ is related to Faulhaber's formula and it holds that
\[
|V_{r,n}|= \sum_{j_{q-1}=0}^{t}\left(\cdots \sum_{j_2=0}^{j_3}\left(\sum_{j_1=0}^{j_2}1\right)\cdots\right) = \frac{t^{q-1}}{(q-1)!} + \Theta(t^{q-2}).
\]
Hence, we obtain
$|V_n| \leq \prod_{i=1}^{M}|V_{r,n}| \leq \left( \frac{t^{q-1}}{q^q} \right)^{M}$. Here, we want to introduce a new notion.  For any subset $A\subseteq V_n$ and any function $f:W\rightarrow\real$, a {\em restricted expectation}
$\check{E}_{A}[f(\hat{v})]$ is defined to be  $\frac{1}{|V_n|}\sum_{v\in V_n} A(v) f(\hat{v})$,
where $A(v)$ is the characteristic function for $A$ (\ie $A(v)=1$ if $v\in A$ and $A(v)=0$ otherwise).
For simplicity, let $\alpha$ denote $(1-\varepsilon)^{\frac{(q-2)M}{2}}q^{\frac{(c-q)M}{2c}}$.
In the rest of this proof, we assume that $q^{MR}\alpha<1$. If we can find
a linear $(M,n)_q$-code $C$ such that, for every $v\in V_n$, there exists a vector $b\in C$ satisfying $\Pre_{\hat{v}}(b)<\varepsilon$,
our assumption $q^{MR}\alpha  <1$ implies that
\[
QU^{const}_{c}(R) \geq \varepsilon = 1-\alpha^{\frac{2}{(q-2)M}} q^{-\frac{c-q}{c(q-2)}} \geq 1 - q^{-\frac{(1+2R)c-q}{(q-2)c}}.
\]
Therefore, the remaining task is to show the existence of a linear code $C$ that satisfies the following condition: for every $v\in V_n$, $|E(\hat{v},\varepsilon)\cap C|\leq c$ holds under the assumption of $q^{MR}\alpha<1$.

Hereafter, we construct $C$ by stages. The notation  $C_i$ expresses a code defined at Stage $i\in[0,n]_{\integer}$ and, in the end of our construction, we set the desired code $C$ to be $C_n$. The key notion for  this construction is the {\em potential function} $S_i$ for $C_i$ defined as
$
S_i = \check{E}_{V_n}[|V_n|^{\frac{1}{c}|E(\hat{v},\epsilon)\cap C_i|}]
$
for each index $i\in[0,n]_{\integer}$. At Stage $0$,  we set $b_0=\bfzero$ and $C_0=span\{b_0\}$. Clearly, for every $v\in V_n$,  we have $|E(\hat{v},\epsilon)\cap C_0|\leq 1$.
Since
$
\Pre_{\hat{v}}(\bfzero) = \frac{1}{M}\braket{v(\bfzero)}{\hat{v}} = \frac{1}{M}\sum_{r}\hat{v}_{r,0},
$
it follows that $\Pre_{\hat{v}}(\bfzero)\geq \epsilon$ iff $\sum_{r\in I_n}\hat{v}_{r,0} \geq \epsilon M$. Next, we consider the set
$
T_t^{(\varepsilon)} = \{v\in V_n\mid \sum_{r\in I_n}\hat{v}_{r,0} \geq \epsilon M t\}.
$
Here, we give a crude estimation to the size of $T_t^{(\varepsilon)}$ as follows. The average value of $\hat{v}_{r,0}$ over all $r\in I_n$ is $t\varepsilon$, and at most a half of them should be at least this value. Hence, each block indexed $r$ contains at most $\frac{(q-1)(1-\varepsilon)^{q-2}}{t}|V_{r,n}|$ possible choices of vectors $v=(v_{r,z})_{z\in\field_q}$. Since there are at most $M!$ possible series $(v_{r,0})_{r\in I_n}$ and $M!\leq t^{(1/2-(q-1)/c)M}$,  $|T_t^{(\varepsilon)}|$ is upper-bounded by
\begin{eqnarray*}
|T_t^{(\varepsilon)}| &\leq& M!\cdot \left( \frac{(q-1)(1-\varepsilon)^{q-2}}{t} \cdot |V_{0,n}| \right)^{\frac{1}{2}M}  |V_{0,n}|^{\frac{1}{2}M} \\
&\leq&  (1-\varepsilon)^{\frac{(q-2)M}{2}}q^{\frac{M}{2}}{t^{-\frac{q-1}{c}M}}{|V_n|}.
\end{eqnarray*}
Note that $v\in T_t^{(\varepsilon)}$ iff $\bfzero\in E(\hat{v},\epsilon)$ iff $|E(\hat{v},\epsilon)\cap C_0|=1$.  Therefore, since $|V_n|^{1/c}\leq \left( \frac{t^{q-1}}{q^q} \right)^{M/c}$,  we can calculate $S_0$ as
\begin{eqnarray*}
S_0 &=& \check{E}_{V_n-T_t^{(\varepsilon)}}[1] + \check{E}_{T_t^{(\varepsilon)}}[|V_n|^{1/c}]
\;\;=\;\;  \frac{|V_n - T_t^{(\varepsilon)}|}{|V_n|} + \frac{|V_n|^{1/c}|T_t^{(\varepsilon)}|}{|V_n|} \\
&\leq& 1+  (1-\varepsilon)^{\frac{(q-2)M}{2}}q^{\frac{(c-q)M}{2c}} \;\;=\;\; 1+\alpha.
\end{eqnarray*}

At Stage $i\geq1$, we choose $b_i$ uniformly at random from $V_n$ so that $b_i$ is linearly independent of $b_1,\ldots,b_{i-1}$.  We define  $C_i=span\{C_{i-1}\cup\{b_i\}\}$. Since $b_i$ is a random variable, so is $C_i$. To complete the construction, we should claim that $|E(\hat{v},\epsilon)\cap C_n| \leq c$ for any $v\in V_n$.
For this purpose, we will define a series $\{\hat{S}_i\}_{0\leq i\leq n}$ of ``average'' values of $S_i$'s, starting with $\hat{S}_0 = S_0$.
Let us consider the conditional expectation $E'_{b_{i+1}}[S_{i+1}\mid S_i = \hat{S}_i]$ over a random choice of $b_{i+1}$ chosen uniformly at random from $V_n-span\{b_1,\ldots,b_i\}$.
Now, we define $\hat{S}_{i+1}$ by $\hat{S}_{i+1} = E'_{b_{i+1}}[S_{i+1}|S_i=\hat{S}_i]$, and we want to show that
$\hat{S}_n\leq 6$.
Let the notation $E_{b_{i+1}}[S_{i+1}\mid S_i = \hat{S}_i]$ be defined similarly, except that $b_{i+1}$ is taken uniformly at random from $V_n$.
Similarly to an argument of Guruswami, H{\aa}stad, Sudan, and Zuckerman \cite{GHSZ02}, it holds that
$E_{b_{i+1}}[S_{i+1}\mid S_i = \hat{S}_i] \leq (\hat{S}_i)^q$.
Since $\frac{|V_n| - q^i}{|V_n|}\cdot E'_{b_{i+1}}[S_{i+1}\mid S_i = \hat{S}_i] \leq  E_{b_{i+1}}[S_{i+1}\mid S_i = \hat{S}_i]$, we conclude that
$E'_{b_{i+1}}[S_{i+1}|S_i=\hat{S}_i] \leq (1-q^{-M+i})^{-1}(\hat{S}_i)^q$ since $|V_n|\geq q^{M}$.
Therefore, when $i=n$, we obtain
\[
\hat{S}_{n} \leq \frac{(\hat{S}_{n-1})^q}{1-q^{-M+n-1}} \leq \frac{(\hat{S}_{0})^{q^n}}{\prod_{i=0}^{n-1}(1-q^{-M+i})^{q^{n-i}}} \leq 2 (\hat{S}_0)^{q^n},
\]
where the last inequality follows from the lower bound
$\prod_{i=0}^{n-1}(1-q^{-M+i})^{q^{n-i}} \geq \frac{1}{2}$.
Since  $q^{MR}\alpha<1$, it follows that
\[
\hat{S}_n \leq 2 (\hat{S}_0)^{q^n} \leq 2\left(1+ \alpha \right)^{q^n} \leq 2\left(1 + 2q^{MR}\alpha  \right) \leq 2(1+2)=6
\]
since $n\leq MR$ and $(1+x)^{m} \leq 1+2mx$ for any  and $m\in\nat^{+}$ and any $x<1/m$.
By the definition of $S_i$, it follows that $\hat{S}_i\geq |V_n|^{-1}|V_n|^{\frac{1}{c}|E(\hat{v},\varepsilon)\cap C_i|}$ for every  $v\in V_n$. In particular, we obtain $|V_n|^{-1}|V_n|^{\frac{1}{c}|E(\hat{v},\epsilon)\cap C_n|} \leq \hat{S}_n\leq 6$, and we therefore conclude that
$
|E(\hat{v},\epsilon)\cap C_n| \leq \left( 1 + \frac{\log 6}{\log|V_n|}\right) c < c +1,
$
as requested.
\end{proofof}

%%%%%%%%%%%%%%%%%
%%%%%%%%%%%%%%%%%

\section*{Acknowledgments}
The author thanks Akinori Kawachi for a discussion on quantum cryptography and Igor Shparlinski for a useful pointer to Reference (Bleichenbacher and Nguyen \cite{BN00}) when the author was preparing the preliminary version of this paper for the conference proceedings of CATS 2007.

%%%%%%%%%%%%%%%%%%%%%%%%%%%%%%%%%%%%
\let\oldbibliography\thebibliography
\bibliographystyle{plain}

%%%%%%%%%%%%%%%%%%%%%%%%%%%%%%%%%%%%%%%%%%%%%
%%%%%%%%%%%%%%%%%%%%%%%%%%%%%%%%%%%%%%%%%%%%%
\end{document}